\documentclass[longauth]{aa} % for the long lists of affiliations 
\usepackage{graphicx}
\usepackage{txfonts}
\def \lsun{\ifmmode{{\rm\ L}_\odot}\else{${\rm\ L}_\odot $}\fi}
\newcommand{\kms}{kms$^{-1}$}                         % Kms-1
\def\iraf{{\sc iraf}}

\def \msun{\ifmmode{{\rm\ M}_\odot}\else{${\rm\ M}_\odot$}\fi}

\def \zsun{\ifmmode{{\rm\ Z}_\odot}\else{${\rm\ Z}_\odot$}\fi}
\newcommand{\ha}{H$\alpha${}}

\newcommand{\hb}{H$\beta${}}

\newcommand{\hii}{H\,{\sc ii}{}}
\newcommand{\cii}{C\,{\sc ii}{}}
\newcommand{\ciii}{C\,{\sc iii}{}}

\newcommand{\oi}{O\,{\sc i}}

\newcommand{\oii}{O\,{\sc ii}}
\newcommand{\caii}{Ca\,{\sc ii}}

\newcommand{\oiii}{[O\,{\sc iii}]}
\newcommand{\feiii}{Fe\,{\sc iii}}

\newcommand{\hei}{He\,{\sc i}}
\newcommand{\nii}{[N\,{\sc ii}]}
\newcommand{\sii}{[S\,{\sc ii}]}

\begin{document}

   \title{A nearby superluminous supernova with a long pre-maximum  `plateau' and strong \cii\ features}

   \author{J.~P. Anderson\inst{1}, P.~J. Pessi\inst{2}, L. Dessart\inst{3}, C. Inserra\inst{4}, 
   D. Hiramatsu\inst{5,6}, K. Taggart\inst{7}, S.~J. Smartt\inst{8}, G. Leloudas\inst{9}, T.-W. Chen\inst{10}, A. M\"oller\inst{11,12}, R. Roy\inst{13}, S. Schulze\inst{14}, D. Perley\inst{7}, J. Selsing\inst{9}, S.~J. Prentice\inst{7,8}, A. Gal-Yam\inst{14}, C.~R. Angus\inst{4},
  I. Arcavi\inst{15,16}\thanks{Einstein Fellow}, C. Ashall\inst{17}, M. Bulla\inst{18}, C. Bray\inst{11}, J. Burke\inst{15,16}, E. Callis\inst{19}, R. Cartier\inst{20}, S.-W. Chang\inst{11,12}, K. Chambers\inst{21}, P. Clark\inst{8}, L. Denneau\inst{21}, M. Dennefeld\inst{22}, H. Flewelling\inst{21}, M. Fraser\inst{19}, L. Galbany\inst{23}, M. Gromadzki\inst{24}, C.~P. Guti\'errez\inst{4}, A. Heinze\inst{21},
  G. Hosseinzadeh\inst{15,16}, D.~A. Howell\inst{15,16}, E.~Y. Hsiao\inst{17}, E. Kankare\inst{8}, 
   Z. Kostrzewa-Rutkowska\inst{25,26}, E. Magnier\inst{21},  K. Maguire\inst{8}, P. Mazzali\inst{7,27}, O. McBrien\inst{8}, C. McCully\inst{15,16}, N. Morrell\inst{28}, T.~B. Lowe\inst{21}, C.~A. Onken\inst{11}, F. Onori\inst{25,26}, M.~M. Phillips\inst{28},
   A. Rest\inst{29,30}, R. Ridden-Harper\inst{11}, A.~J. Ruiter\inst{31,32}, D.~J. Sand\inst{33}, K.~W. Smith\inst{8}, M. Smith\inst{4}, B. Stalder\inst{34}, 
   M.~D. Stritzinger\inst{35}, M. Sullivan\inst{4}, J.~L. Tonry\inst{21}, B.~E. Tucker\inst{12,32,36,37}, S. Valenti\inst{38}, R. Wainscoat\inst{21}, C.~Z. Waters\inst{21}, C. Wolf\inst{11,12}, D. Young\inst{8}}

   \institute{European Southern Observatory, Alonso de C\'ordova 3107, Casilla 19, Santiago, Chile. \email{janderso@eso.org}
         \and Instituto de Astrofísica de La Plata (IALP), CONICET, Argentina; Facultad de Ciencias Astronómicas y Geofísicas, Universidad Nacional de La Plata, Paseo del Bosque, B1900FWA, La Plata, Argentina
         \and Unidad Mixta Internacional Franco-Chilena de Astronom\'ia (CNRS UMI 3386), Departamento de Astronom\'ia, Universidad de Chile, Camino El Observatorio 1515, Las Condes, Santiago, Chile
         \and School of Physics and Astronomy, University of Southampton, Southampton, UK
         \and Las Cumbres Observatory, 6740 Cortona Drive, Suite 102, Goleta, CA 93117-5575, USA
         \and Department of Physics, University of California, Santa Barbara, CA 93106-9530, USA
         \and Astrophysics Research Institute, Liverpool John Moores University, 146 Brownlow Hill, Liverpool L3 5RF, UK
         \and Astrophysics Research Centre, School of Mathematics and Physics, Queens University Belfast, Belfast BT7 1NN, UK
         \and Dark Cosmology Centre, Niels Bohr Institute, University of Copenhagen, Juliane Maries vej 30, 2100 Copenhagen, Denmark
         \and Max-Planck-Institut für Extraterrestrische Physik, Giessenbachstraße 1, 85748, Garching, Germany
         \and Research School of Astronomy and Astrophysics, Australian National University, Canberra, ACT 2611, Australia.
 \and ARC Centre of Excellence for All-sky Astrophysics (CAASTRO), Australia         \and Inter-University Centre for Astronomy and Astrophysics, Ganeshkhind, Pune - 411007, Maharashtra, India
         \and Department of Particle Physics and Astrophysics, Weizmann Institute of Science, Rehovot, 7610001, Israel
         \and Las Cumbres Observatory, 6740 Cortona Dr. Suite 102, Goleta, CA, 93117-5575, USA
         \and University of California, Santa Barbara, Department of Physics, Santa Barbara, CA, 93106-9530, USA
         \and Department of Physics, Florida State University, Tallahassee, FL 32306, USA
         \and Oskar Klein Centre, Department of Physics, Stockholm University, SE 106 91 Stockholm, Sweden
         \and School of Physics, O’Brien Centre for Science North, University College Dublin, Belfield, Dublin 4, Ireland
         \and Cerro Tololo Inter-American Observatory, National Optical Astronomy Observatory, Casilla 603, La Serena, Chile 
           \and Institute for Astronomy, University of Hawaii, 2680 Woodlawn Drive, Honolulu, HI 96822
           \and Institut d'Astrophysique de Paris, CNRS, and Universite Pierre et Marie Curie, 98 bis Boulevard Arago, F-75014 Paris, France
         \and PITT PACC, Department of Physics and Astronomy, University of Pittsburgh, Pittsburgh, PA 15260, USA
         \and Warsaw University Astronomical Observatory, Al. Ujazdowskie 4, 00-478
Warszawa, Poland
         \and SRON Netherlands Institute for Space Research, Sorbonnelaan 2, 3584 CA Utrecht, the Netherlands
         \and Department of Astrophysics/IMAPP, Radboud University Nijmegen, P.O. Box 9010, 6500 GL Nijmegen, the Netherlands
         \and Las Campanas Observatory, Carnegie Observatories, Casilla 601, La Serena, Chile
         \and Max-Planck-Institut f{\"u}r Astrophysik, Karl-Schwarzschild-Str. 1, D-85748 Garching, Germany
         \and Space Telescope Science Institute, 3700 San Martin Drive, Baltimore, MD 21218, USA
         \and Department of Physics and Astronomy, Johns Hopkins University, Baltimore, MD 21218, USA
         \and ARC Future Fellow, School of Physical, Environmental and Mathematical Sciences, University of New South Wales, Australian Defence Force Academy, Canberra, ACT 2600, Australia
         \and Research School of Astronomy and Astrophysics, Mt Stromlo Observatory, Australian National University, Canberra, ACT 2611, Australia
         \and Department of Astronomy/Steward Observatory, 933 North Cherry Avenue, Rm. N204, Tucson, AZ 85721-0065, USA
         \and LSST, 950 N Cherry Ave, Tucson, AZ 85719
         \and Department of Physics and Astronomy, Aarhus University, Ny Munkegade 120, DK-8000 Aarhus C, Denmark
         \and The National Centre for the Public Awareness of Science, Australian National University, Canberra, ACT 2611, Australia
         \and ARC Centre of Excellence for All-sky Astrophysics in 3 Dimensions (ASTRO 3D), Australia
         \and Department of Physics, University of California, Davis, CA 95616, USA
             }
\titlerunning{A nearby SLSN with strong \cii\ features}
\authorrunning{Anderson et al.}
   \date{}
   
% \abstract{}{}{}{}{} 
% 5 {} token are mandatory
 
  \abstract
  % context heading (optional)
  % {} leave it empty if necessary  
   {Super-luminous supernovae (SLSNe) are rare events defined as being significantly more luminous than 
   normal terminal stellar explosions. The source of the additional power needed to achieve such luminosities is still unclear.
   Discoveries in the local Universe (i.e. $z<0.1$) are 
   scarce, but afford dense multi-wavelength observations. Additional low-redshift objects are
   therefore extremely valuable.
   }
  % aims heading (mandatory)
   {We present early-time observations of the type I SLSN ASASSN-18km/SN~2018bsz.
   These data are used to characterise the event and compare to 
   literature SLSNe and spectral models. Host galaxy properties are also analysed.}
  % methods heading (mandatory)
   {Optical and near-IR photometry and spectroscopy were analysed. Early-time ATLAS photometry was used to constrain the rising light curve.  We identified a number of spectral features in optical-wavelength spectra and track their time evolution. Finally, we
   used archival host galaxy photometry together with \hii\ region spectra to constrain the host environment.}
  % results heading (mandatory)
   {ASASSN-18km/SN~2018bsz is found to be a type I SLSN in a galaxy at a redshift of 0.0267 (111\,Mpc), making it the lowest-redshift event discovered to date. Strong \cii\ lines are identified in the spectra. Spectral models produced by exploding a Wolf-Rayet progenitor and injecting a magnetar power source are shown to be qualitatively similar to ASASSN-18km/SN~2018bsz, contrary to most SLSNe-I that display weak or non-existent \cii\ lines. 
   ASASSN-18km/SN~2018bsz displays a long, slowly rising, red `plateau' of $>$26 days, before a steeper, faster rise to maximum. 
   The host has an absolute magnitude of --19.8 mag ($r$), a mass of M$_{*}$ = 1.5$^{+0.08}_{-0.33}$ $\times$10$^{9}$ M$_{\odot}$ , and a star formation rate of = 0.50$^{+2.22}_{-0.19}$ M$_{\odot}$ yr$^{-1}$. A nearby \hii\ region has an oxygen abundance (O3N2) of 8.31$\pm0.01$ dex.}
  % conclusions heading (optional), leave it empty if necessary 
   {}


   \maketitle
%
%________________________________________________________________

\section{Introduction}
SLSNe are a class of transients that have exceptionally high luminosities (\citealt{gal09b,pas10,cho11,qui11,gal12a}, see \citealt{how17} for a review).
Historically, they were defined as being brighter than --21 mag at optical bands. However, such a specific limit is somewhat arbitrary and as larger samples have been assembled classification is now based on morphological similarity in addition to a brightness limit \citep{ins18b,qui18,dec18}.\\
\indent First identified more than a decade ago, there now exist samples of several tens of well-observed objects (\citealt{nic14,ins18a,lun18,ins18b,qui18,dec18}). However, given their low rates per unit volume (\citealt{qui13,mcc15,pra17}), nearby events are scarce with most objects discovered at redshifts higher than 0.1. This can limit the observability and observed wavelength regime of SLSNe to only the brightest phases, limiting the physics that can be extracted from observations.\\ 
\indent SLSNe are classified into several spectroscopic groups \citep{gal12a}. Broadly, they are separated into
hydrogen-rich events, SLSNe-II (that are dominated by narrow-line SLSNe; for example SN~2006gy, \citealt{smi07,ofe07}, but also contain extremely bright broad-line events such as SN~2008es, \citealt{gez09,mil09}, and the small sample presented in \citealt{ins18a}), and the more
numerous (in number of discovered events) hydrogen-poor SLSNe-I (see \citealt{qui18} and references therein). These latter events show some spectroscopic similarity to type Ic SNe (SNe~Ic) after maximum light
(e.g. \citealt{pas10,liu17}), and modelling has suggested their spectra can be produced by the explosion of massive carbon-oxygen cores \citep{des12,how13}.
SLSNe-I display considerable diversity in their photometric and spectral evolution \citep{qui18,dec18}, and further sub-classification into fast and slowly evolving
events has also been discussed \citep{gal12a,nic15b,ins17,ins18b,qui18}.\\
\indent Several physical mechanisms have been proposed as the source of the additional power required by SLSNe (see review by \citealt{mor18a}). An increase of radioactive material can significantly increase the available energy of an event, and very large synthesised $^{56}$Ni masses are an outcome of Pair-Instability SNe (PISNe, see \citealt{heg02}; \citealt{gal09b} and references therein). Interaction of a SN ejecta with dense circumstellar material (CSM) has been proposed, not only for those SLSNe showing narrow hydrogen lines but also for SLSNe in general \citep{che11,gin12,des15,sor16}. Accretion of material onto a central compact object post core-collapse can also produce an additional energy source \citep{dex13,mor18b}. Finally, trapping the power output from a fast rotating neutron star: a magnetar, is regularly discussed as a viable option \citep{kas10,woo10}.
While different prescriptions of the above scenarios have been shown to fit some aspects of SLSNe, a clear consensus on the dominating power source that explains the origin of SLSNe is still lacking. Additional nearby events where one can obtain high-cadence multi-wavelength observations can be highly constraining for our
understanding of SLSNe.\\

\indent Here, we analyse the nearby SN, ASASSN-18km/SN~2018bsz, which we henceforth refer to as SN~2018bsz (the IAU designated name). While initial reports classified this event as a type II SN (SN~II, see \citealt{hir18,cla18}), a reclassification to a SLSN-I was later announced (\citealt{and18}). A host redshift of 0.0267 \citep{jon09b} makes SN~2018bsz the closest SLSN-I discovered to date, enabling a detailed multi-wavelength observational campaign.
SN~2018bsz exploded in the galaxy 2MASX J16093905-3203443 and is affected 
by Milky Way line of sight extinction of E($B$-$V$)=0.214 mag \citep{Sch11} assuming a
\cite{fit99} reddening law and an R$_{V}$ of 3.1.
Throughout the paper we assume an $H_0$ of 73 km s$^{-1}$ Mpc$^{-1}$ and
a standard cosmology ($\Omega_{m}$ = 0.27, $\Omega_\lambda$ = 0.73; \citealt{spe07}). \\
\indent We present UV, optical and near-IR photometry and spectroscopy of SN~2018bsz ranging from 50 days before to a week past maximum optical light. These data are compared to other well-observed events, and are used to characterise the overall behaviour of 
SN~2018bsz in the context of our current understanding of SLSNe. In the next section we summarise the discovery, classification and reclassification of SN~2018bsz. In Section 3 we present our spectroscopy. Section 4 presents the light curve of SN~2018bsz and Section 5 compares our spectra to those produced by radiative transfer models. In Section 6 we analyse the SN environment properties. We finish in Section 7 with our conclusions.

\begin{figure}
\centering
\includegraphics[width=\columnwidth]{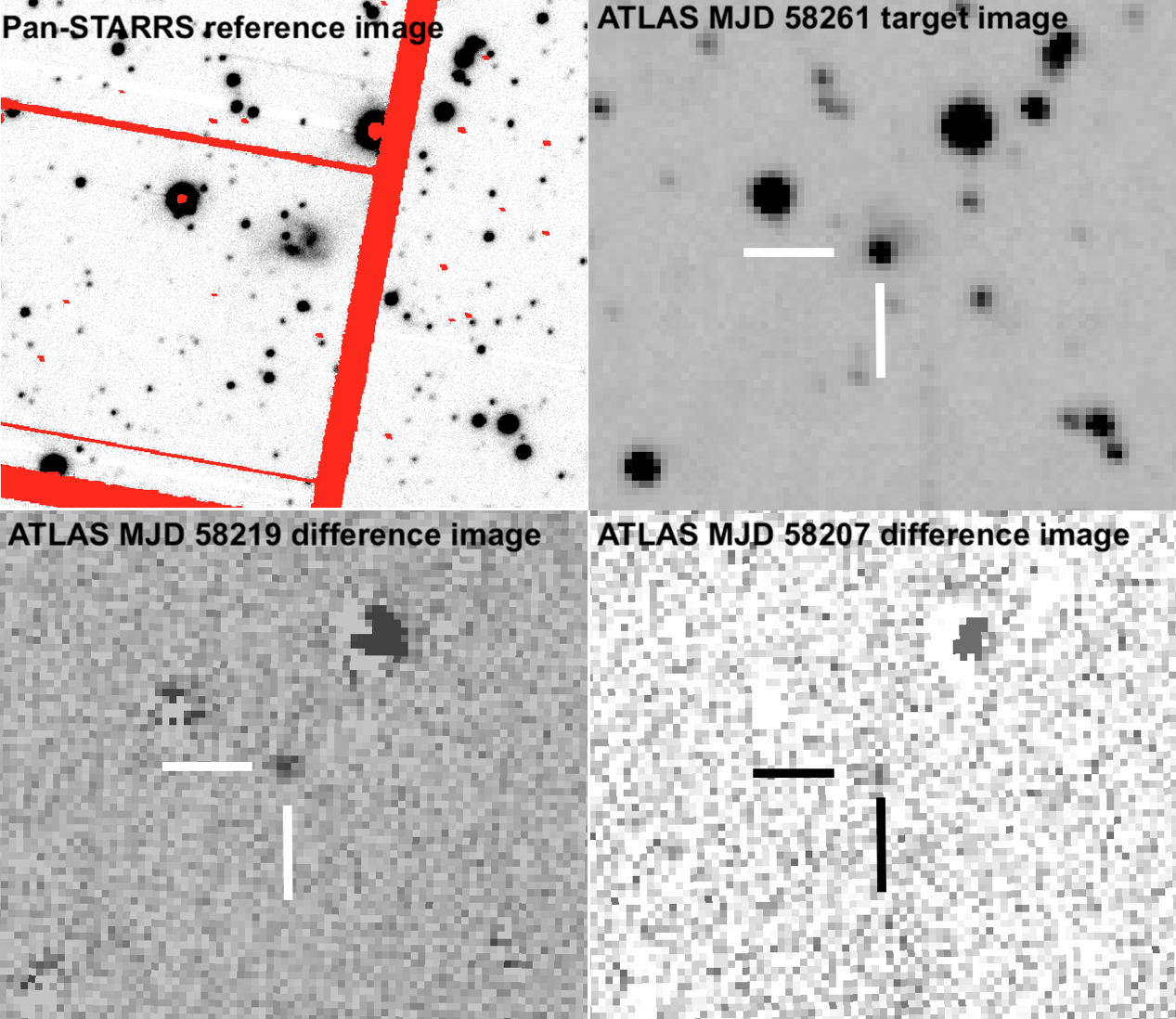}
\caption{\textit{Top left:} Pan-STARRS reference $i$-band image of the host of SN~2018bsz (the red pixels are bad pixels or chip gaps). \textit{Top right:} ATLAS target image from 2018 May 23. \textit{Bottom left:} ATLAS difference image obtained on 2018 April 11, showing a clear detection of SN~2018bsz. \textit{Bottom right:} ATLAS difference image obtained on 2018 March 30 showing the 4-sigma detection.}
\label{diffimg}
\end{figure}

\begin{figure*}
\centering
\includegraphics[width=19cm]{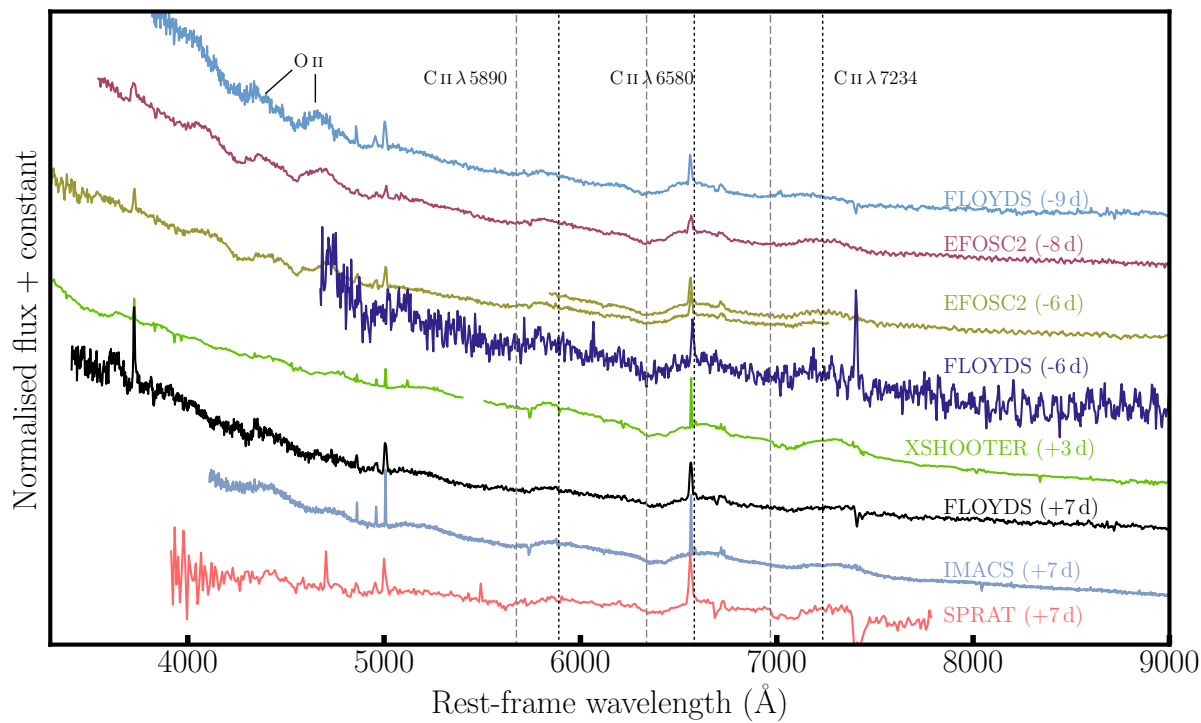}
\caption{Optical wavelength spectral sequence of SN~2018bsz: from nine days before to seven days after maximum light (MJD = 58267.5 in $r$). 
Dotted vertical black lines indicate the rest wavelength of three \cii\ lines, while the dashed vertical grey lines indicate the same \cii\ lines at a velocity of --11,000\,\kms. Two additional lines are clearly observed between 4000 and 5000\,\AA\ that are most likely associated with \oii. Narrow emission is visible due to an underlying host \hii\ region (see e.g. \ha, \hb) and is not associated with the transient itself. Additional line identifications are presented
in Fig.~\ref{bestseq}.
Spectra have been corrected for line of sight extinction, both in our own Milky Way and from within
the host galaxy of SN~2018bsz. We do not plot two spectra obtained with  SPRAT at +3 days and with FLOYDS at +3 days, due to their low S/N (but their details are listed in Table~\ref{tabspec} and the spectra will be released). The X-Shooter spectra have been binned to 6\,\AA\ in wavelength.}
\label{specseq}
\end{figure*}

\section{Discovery, classification and reclassification}
SN2018bsz was discovered \citep{sta18,bri18} by the All Sky Automated Survey for
SuperNovae\footnote{http://www.astronomy.ohio-state.edu/$\sim$~assassin/index.shtml}, ASAS-SN \citep{sha14}
as ASASSN-18km on 2018 May 17 (MJD = 58255.97) at a $g$-band magnitude of 17.3. 
The object was independently detected as a transient 
by the ATLAS survey\footnote{http://fallingstar.com/}
\citep{ton18}  
as ATLAS18pny
on 2018 May 21 (MJD 58259.66), with improved coordinates 
of R.A. = 16:09:39.11  Decl.=-32:03:45.63. 
Automated forced photometry 
on the ATLAS difference images produced detections ($>5\sigma$
significance) back to 2018 April 11 (MJD = 58219.53). 
We manually inspected the two preceding nights of observations and found that
stacking the four 30\,s frames on the night of 2018 Mar 30 
(MJD = 58207.46) produced a 4.4$\sigma$ detection at
$o=18.73\pm0.25$ ($o$ refers to the ATLAS `orange' band that has a wavelength range of 5600 to 8200\,\AA). The stacked ATLAS images a week prior to maximum light, plus the two difference images for the detections found through forced photometry are displayed in Fig.~\ref{diffimg}, where we also present a Pan-STARRS \citep{cha16} pre-explosion $i$-band image of the host.
We found a non-detection on the stack 
of 4$\times$30\,s cyan filter (ATLAS $c$ band, wavelength range 4200 to 6500\,\AA) images taken on
MJD=58193.57, and one single $c$-band image taken on 58197.56. This latter date is taken
as the last non-detection prior to discovery (limiting AB magnitude of 20.32 mag).\\
\indent The explosion date is then estimated as the mid-point between the epoch of last non detection and that of discovery, with the error being half the difference in time between the points. We therefore adopt an estimated explosion epoch for SN~2018bsz of 2018 March 25, MJD = 58202.5$\pm$5. It is important to note that we cannot rule out a significantly earlier epoch with SN~2018bsz evolving at magnitudes below our photometric limits. This would require a steeper initial rise before the `long duration plateau' (the latter is characterised in Section 4). However, the value of the true explosion date and the nature of any prior unseen light-curve evolution does not strongly affect our main results and conclusions.\\
\indent Initial spectral classifications were made on 2018 May 20 \citep{hir18,cla18} by both the Las Cumbres Global Supernova Project, and ePESSTO\footnote{http://www.pessto.org}, the extended Public ESO Spectroscopic Survey for Transient Objects (\citealt{sma15b}). Both reports  concluded SN~2018bsz was a young SN~II due to a strong P-Cygni spectral line profile at a wavelength consistent with \ha\ (see spectra in Fig.~\ref{specseq}).
At the redshift of the host galaxy (0.0267), this made SN~2018bsz a bright SN~II. However, further inspection of the spectral features 
cast doubt on this initial classification.\\
\indent A clear bump is visible to the red of \ha, that 
is not generally observed in SNe~II. \hb\ is usually stronger in absorption than \ha\ in early-time SN~II spectra (see e.g. \citealt{gut17a}), however there is no evidence for this line in any of the spectra of SN~2018bsz (see Fig.~\ref{specseq}). 
(In SLSNe-II \ha\ is also always accompanied by strong \hb: \citealt{mil09,gez09,ter17,ins18b}.)
There are a number of spectral lines observed between 4000 and 5000\,\AA\ that are hard to interpret within
a SN~II classification. Comparing the classification spectrum to SLSN spectral models of Dessart (in preparation, see Section 5), suggested that the strong P-Cygni feature at around 6500\,\AA\ could be \cii\,$\lambda$\,6580, with additional \cii\ features also identified.
The Supernova Identification (SNID, \citealt{blo07}) software package was used
and good matches were found between SN~2018bsz (using the earliest spectrum; nine days before maximum) and several SLSNe, including PTF12dam at 15 days before maximum and PTF09atu at 20 days before maximum.
In addition, SN~2018bsz continued to rise to an absolute magnitude of --20.5 mag (in $B$, after correction for Milky-Way and host-galaxy extinction, see Section 3.2). This brightness falls at the low end of the distribution of SLSN-I luminosities \citep{dec18,ins18b} but is still much brighter than canonical SNe~Ic (and core-collapse events in general; \citealt{ric14}). These properties led to a reclassification of SN~2018bsz to a SLSN-I \citep{and18}.

\section{Spectral properties of SN~2018bsz}

\subsection{Spectral observations}
\begin{figure}
\centering
\includegraphics[width=\columnwidth]{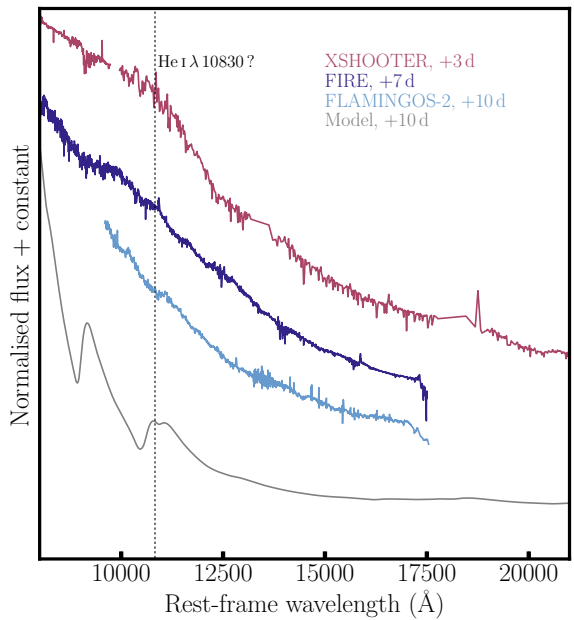}
\caption{Near-IR spectral sequence of SN~2018bsz. We also plot the model of Dessart (in preparation) that is discussed in Section 5. The rest wavelength of He\,{\sc i}\,$\lambda$\,10830 is indicated.}
\label{nearIRseq}
\end{figure}

In Figure~\ref{specseq} we present the optical-wavelength spectral sequence of SN~2018bsz.
Details of this sequence are listed in Table~\ref{tabspec}. These data were obtained and reduced
through standard procedures. FLOYDS spectra from the Las Cumbres Observatory \citep{bro13} Global Supernova Project were reduced as in \cite{val14}, while
EFOSC2 (\citealt{buz84}, mounted on the NTT) spectra were reduced using a custom built pipeline for the PESSTO project \citep{sma15b}.
Spectra obtained by the Spectrograph for the Rapid Acquisition of Transients (SPRAT, \citealt{pia14}) on the Liverpool Telescope \citep{ste04} were reduced using the standard SPRAT pipeline.
IMACS (mounted on the Magellan telescope) was used in its short camera configuration (f/2) with a 0.9$\arcsec$ slit and the 300 l/mm blue grism. Reductions were achieved using standard \iraf\footnote{\iraf\ is distributed 
by the National Optical Astronomy Observatory, which is operated by the 
Association of Universities for Research in Astronomy (AURA) under 
cooperative agreement with the National Science Foundation.} routines.
Near-UV, optical and near-IR spectroscopic observations carried out with X-Shooter (\citealt{ver11}, on the VLT) were obtained in a single 900 s exposure, using $1.0\times11\arcsec$, $0.9\times11\arcsec$, and $0.9\times11\arcsec$ slits for the UVB, VIS, and NIR arms respectively. The spectra were reduced as outlined in \citet{sel18}, including: recalibrating the wavelength solution, correcting for slit losses, and correcting for telluric absorption.\\
\indent Near-IR wavelength spectra were obtained at three epochs and are presented in
Fig.~\ref{nearIRseq} (see Table~\ref{tabspec} for details). 
The FLAMINGOS-2 (\citealt{eik08}, on the Gemini telescope) spectrum was taken in longslit mode with the $JH$ grism and filter in place, with a slit width of 0.72", yielding a wavelength range of 1.0--1.8 $\mu$m and $R$$\sim$1000. The data were obtained at the parallactic angle with a standard ABBA pattern for sky subtraction. The total exposure time was 12$\times$120\,s. The data were reduced in a standard way using the FLAMINGOS-2 \texttt{PyRAF} package provided by the Gemini observatory. Telluric corrections and flux calibration were determined with an A0V star observed adjacent in time to the SN~2018bsz data, using the telluric correction methodology of \citet{vac03}. The FIRE (mounted on the Magellan telescope) spectrum was taken with the high throughput long slit mode with 0.6" slit width. A total of 4 sets of ABBA were obtained, with per frame exposure time of 126.8 s, making a total on-target exposure time of 2028.8 s. The A0V star, HD148733, was used as the telluric and flux standard, and standard reduction routines were followed. All spectral observations will be made public via WISeREP \citep{yar12}\footnote{https://wiserep.weizmann.ac.il/}.

\begin{figure}
\includegraphics[width=\columnwidth]{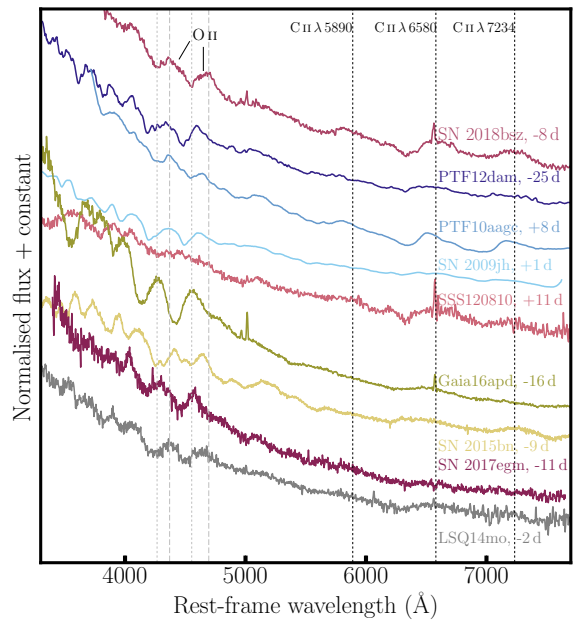}
\caption{Spectral comparison between SN~2018bsz at --9 day (from maximum light) and SLSNe of the best SNID matches together with three well observed, nearby
literature events, and a well-observed fast-evolving SLSN-I. The rest wavelength of the three most prominent \cii\ lines are shown by dotted black lines, while the peaks and troughs of the w-shaped \oii\ feature for SN~2018bsz are indicated by dashed and dotted grey lines respectively. The latter enables a direct comparison to the peaks and troughs found in other SLSNe-I.
\tiny{\textit{References:} PTF12dam \citep{qui18}; PTF10aagc \citep{qui18}; SN~2009jh \citep{qui18}, these first three are smoothed spectra;
SSS120810 (full name SSS120810-231802-560926, \citealt{nic14}), Gaia16apd \citep{kan17}; SN~2015bn \citep{nic16b}; SN~2017egm \citep{bos18}; LSQ14mo \citep{che17c}.}}
\label{spesnid}
\end{figure}

\subsection{Spectral-line identification, time evolution, and comparison to other SLSNe}
\begin{figure}
\centering
\includegraphics[width=\columnwidth]{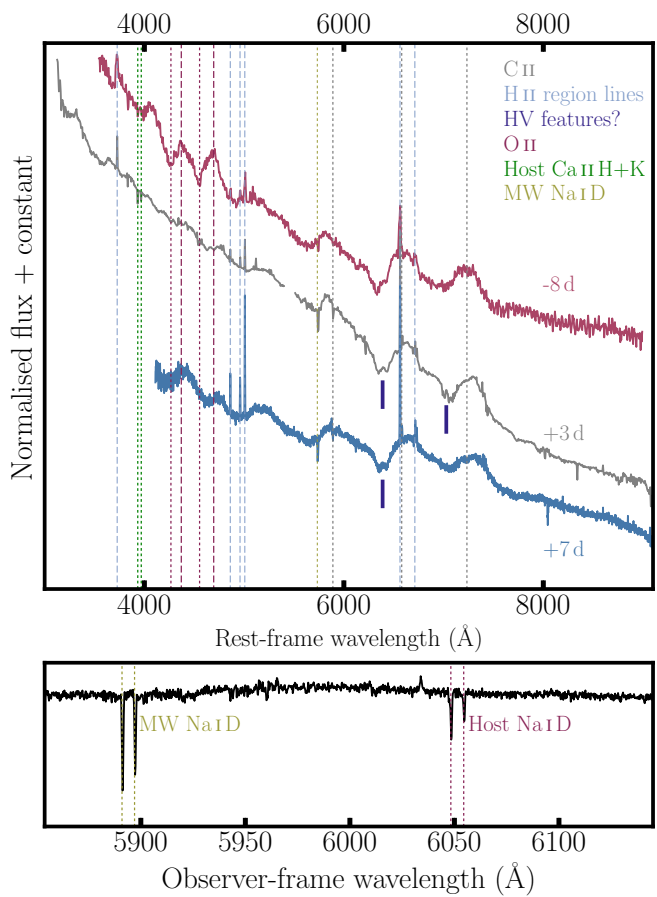}
\caption{\textit{Top:} SN~2018bsz optical-wavelength spectra obtained at eight days prior to maximum light, three days after maximum light, and one week post maximum. 
The positions of host \hii\ region emission lines (not associated with the SN) are indicated as dashed blue vertical lines, while narrow absorption from Milky-Way Na\,{\sc i}\,D 
absorption is shown as a dashed yellow line (and a zoom also showing such absorption within the host is presented in the bottom panel). \cii\ line rest wavelengths are indicated by grey vertical lines, while the position of both the peaks and the troughs of the \oii\ features is shown in red. We also identify relatively narrow emission features in the absorption troughs of \cii, that we label as `High Velocity, HV' features.
Narrow calcium H\,+\,K absorption from within the host galaxy is also observed.  
\textit{Bottom:} a zoom of the spectral region between 5800\,\AA\ and 6150\,\AA\ as seen in the X-Shooter spectrum (three days post maximum). The latter shows the clear detection of sodium absorption within the host galaxy of SN~2018bsz, together with the strong
MW sodium absorption.}
\label{bestseq}
\end{figure}

A number of strong features are observed in the optical spectra. Specifically, we identify 
\cii\ lines at wavelengths of $\sim$5890\,\AA, $\sim$6580\,\AA\ and $\sim$7234\,\AA. 
These \cii\ lines show consistent blue-shifted absorption troughs at around 11000\,\kms\ (Fig.~\ref{specseq}) in the spectra obtained around a week before maximum light (alternatively, around 55 days post explosion), which are typical of SLSNe-I near to peak \citep{gal18}. Such strong carbon lines have been predicted by spectral models 
(\citealt{des12,maz16}), but have rarely been detected at such strength (but see \citealt{nic16b}, for identification of \cii\ in SN~2015bn, together with \citealt{yan17} where a number of \cii\ lines are detected and characterised in Gaia16apd).\\
\indent SLSN-I spectra are usually characterised as having relatively strong spectral features between 3500 and 5000\,\AA\ associated with \oii\ and formed by many tens of overlapping
lines (see modelling of \citealt{maz16} for lists of line identifications). The characteristic `w-shape' produced by these \oii\ lines is seen in the early-time spectra of SN~2018bsz, however the whole feature appears shifted to redder wavelengths than usual (Fig.~\ref{spesnid}, although they do nicely align with PTF10aagc as will be discussed below). 
The features usually
associated with \oii\ are formed by many tens of overlapping lines.  
The absorption minima are formed at wavelengths where these multiple \oii\ lines are blue shifted to, with peaks seen at spaces where \oii\ contributions are not present \citep{gal18}.
Fig.~\ref{spesnid} clearly shows that the wavelengths of the peaks in these regions for SN~2018bsz (dashed vertical grey lines) align with the wavelengths of troughs \textit{(absorption)} for the two well-observed nearby SLSNe-I: Gaia16apd and SN~2017egm.\\ 
\indent The reasons for these clear offsets are not obvious. Assuming the first absorption trough in SN~2018bsz and Gaia16apd as one moves blue ward of \hb\ is produced by the same feature, then comparing the wavelength of these absorption troughs one obtains a velocity difference of more than 10,000\,\kms, which is too large to be explained through ejecta-velocity differences.
It is possible that 
a change in the morphology of the spectrum in this wavelength region (between SNe) may be produced through differences in ejecta density profiles. In addition, differences could be caused by overlapping lines such as \feiii. Finally, identifying these features as the same \oii\ lines in different SLSNe-I may be in error. We leave a more in-depth study of this latter possibility for future work.\\ 
\indent The near-IR spectra (Fig.~\ref{nearIRseq}) are almost
completely featureless. In Fig.~\ref{nearIRseq} we indicate the rest wavelength of 
\hei\,$\lambda$\,10830, and there is a small feature in the observed spectra slightly to the red that can most easily be seen in the FLAMINGOS-2 spectrum. We do not claim a clear detection of any line here (whether it be \hei\,$\lambda$\,10830 or something else), but we do note that \hei\,$\lambda$\,10830 is relatively strong in the spectral model (displayed in Fig.~\ref{nearIRseq} and discussed in more detail below), possibly supporting the line detection and identification. Spectral-line detections at such wavelengths in SLSNe-I have previously been claimed for SN~2012il by \cite{ins13b} and in SN~2015bn by \cite{nic16b}. However, similarly to SN~2018bsz, an identification with \hei\,$\lambda$\,10830 is still not clear.\\
\indent In Fig.~\ref{bestseq} we plot three of the highest S/N optical-wavelength spectra that we have obtained of SN~2018bsz that also cover the full range of epochs at which SN~2018bsz has been observed. The three clearly detected \cii\ lines discussed above ($\lambda$\,5890, $\lambda$\,6580 and $\lambda$\,7234) do not appear to change visually in shape or strength over the 16 day time range. Measuring line velocities from the minimum flux of absorption (by fitting a Gaussian to the line using \iraf), we find that on average absorption velocities go down by 3000\,\kms\ from eight days before to three days after maximum light, to around 8000\,\kms. There is little evolution between four and seven days post maximum.\\ 
\indent Defining the wavelength of minimum absorption is complicated by the presence of small, narrow emission features directly within the absorption trough of \cii\,$\lambda$\,6580 and possibly \cii\,$\lambda$\,7234, as indicated by the dark blue lines in Fig.~\ref{bestseq}. These features are most prominent in the X-Shooter (+3 days) and 
IMACS (+7 days) spectra, but can also be identified in the EFOSC2 spectrum eight days before maximum. Analysing this feature in the X-Shooter spectrum (that of the highest spectral resolution), we measure a FWHM of the emission of 2300\,\kms, and the peak of the emission falls at a rest wavelength of 6400\,\AA. We tentatively identify these as High-Velocity (HV) features of \cii\ being formed in a detached shell of material at almost 9000\,\kms\ (as compared to the bulk ejecta velocities of around 8000\,\kms\ at these epochs).\\
\indent In the bottom panel of Fig.~\ref{bestseq} we present the X-Shooter spectrum in the wavelength range 5800 to 6510\,\AA\ in order to show the wavelength region of absorption from the sodium doublet (Na\,{\sc i}\,D). Absorption is clearly observed at both rest-frame wavelength of the doublet ($\lambda\lambda$\,5896,5890) and at the redshift of SN~2018bsz (the absorption is also clearly seen in the IMACS spectrum). Measuring the equivalent widths of each line, for the Milky-Way absorption we estimate a total
Na\,{\sc i}\,D EW of 0.91\,\AA. Using this value, together with the relation from \cite{poz12} we obtain a E($B$-$V$)=0.164\,mag that is somewhat lower but still consistent with the value quoted above taken from \cite{Sch11} (given the dispersion on the \citealt{poz12} relation together with our measurement errors). 
The detection of Na\,{\sc i}\,D absorption within the host of SN~2018bsz is the first time (to our knowledge) such absorption has been identified in spectra of SLSNe-I. Host galaxy extinction is usually assumed to be negligible in SLSN hosts \citep{lel15,nic15b,dec18}. 
At the redshift of the host galaxy we measure a total EW (summing the two lines) of 0.40\,\AA.
This translates to a E($B$-$V$)=0.041\,mag along the line of sight to SN~2018bsz within the host galaxy.
We use this value to further correct spectra and photometry for host galaxy line-of-sight extinction. In addition to Na\,{\sc i}\,D absorption, we also observe \caii\,H and K absorption (at 3966,3934\,\AA) at the redshift of the host. A deeper analysis of these narrow interstellar (or circumstellar) lines and a search for possible time variability, will be the focus of future work.\\
\indent As outlined above, our reclassification of SN~2018bsz was aided by comparison to spectral templates using SNID \citep{blo07}, together with the comparison to spectral models (Dessart in preparation). In order to find similar events we searched the Open
Supernova Catalog \citep{gui17c} for all SLSN-I with more than three spectra of reasonable S/N (in order to have various phases of the template SNe covered). These spectra were then converted into templates for SNID, and a comparison was made between SN~2018bsz and other SLSN spectra (together with the full range of other SN spectral templates: those originally available in SNID plus those from \citealt{liu16,liu17} and \citealt{gut17a}). The best matches (according to SNID but then confirmed visually\footnote{The top three SNID matches were all SLSNe, the next matches (to non-SLSNe) were discarded visually.}) are compared to SN~2018bsz in Fig.~\ref{spesnid}, together with those from additional three well-observed nearby SLSNe-I.
Reasonable matches were found to: SN~2009jh; PTF12dam; and SSS120810 (references for these SNe are listed in the figure caption). The best match was found with PTF10aagc. PTF10aagc clearly shows spectral features at 
similar wavelengths to SN~2018bsz, specifically \cii\,$\lambda$5890 $\lambda$\,6580 and $\lambda$\,7234, however the emission peaks appear to be consistently blue shifted. PTF10aagc also shows similarities to SN~2018bsz in the wavelength position of the features at around 4500\,\AA. 
Similarly to SN~2018bsz (see below), PTF10aagc was relatively low-luminosity for a SLSN, with a peak rest-frame $g$-band brightness of --20.1 mag \citep{dec18}. While PTF10aagc lacks pre-maximum data, its post-peak observations showed two interesting properties. First, PTF10aggc evolved extremely rapidly, being one of the fastest events in the sample of \cite{dec18}. Here we do  not discuss in any detail the evolution of SN~2018bsz after maximum light, however there are already some indications that it will fall into the `fast' evolving group of SLSNe-I. Second, in spectra obtained several months after maximum PTF10aagc developed clear hydrogen features, with convincing detections of both \ha\ and \hb\ \citep{yan15,qui18}. We await the post-peak spectral evolution of SN~2018bsz to see if it reveals similar features.\\
\indent While a number of SLSNe-I were found to match the early-time spectra of SN~2018bsz, the three nearby events: SN~2015bn \citep{nic16b},  Gaia16apd \citep{kan17}, and SN~2017egm \citep{bos18}, which we also compare to in Fig.~\ref{spesnid} show similarities but clear differences. Indeed, Gaia16apd can be found to match spectra of SN~2018bsz, but only if redshift is allowed to vary enabling the 4000--5000\,\AA\ features to match (SNID gives matches with SN~2018bsz to many SLSNe-I but at significantly different redshifts to its host). This again highlights the issue discussed above: the \oii\ identifications in SN~2018bsz are quite distinct from those in other nearby SLSNe.
Gaia16apd was as an intermediate object in terms of its light-curve evolution; in between those of fast and slow evolving SLSNe-I, while being relativity bright at a $g$-band absolute magnitude of --21.8 mag \citep{kan17}. As shown in Fig.~\ref{spesnid}, Gaia16apd displayed strong \oii\ features with peaks and troughs at significantly different rest wavelengths to SN~2018bsz. A number of \cii\ features were identified in Gaia 16apd by \cite{yan17}, but weaker than in SN~2018bsz.\\
\indent SN~2015bn was intrinsically brighter than the general SLSN-I population (at \textit{M$_{U}$} of around --23.1 mag), faded slowly, and displayed
significant variability in its decline post-peak \citep{nic16b}. \cii\,$\lambda$\,7234 was identified \citep{nic16b} and a hump can clearly be seen in Fig.~\ref{spesnid} at these wavelengths. However, no significant \cii\,$\lambda$\,6580 was observed. SN~2015bn showed relatively strong \oi\,$\lambda$\,7774 absorption that is not seen in SN~2018bsz.\\
\indent Similar to SN~2018bsz, SN~2017egm was relatively faint for a SLSN-I \citep{bos18}. \cii\ was tentatively identified by \cite{bos18} in the earliest spectrum of SN~2017egm, and clear absorption related to \cii\,$\lambda$\,6580 developed around maximum light, however none of the \cii\ features were as strong as seen in SN~2018bsz. 
LSQ14mo is also displayed in Fig.~\ref{spesnid} as an example of a well-observed fast-evolving SLSN-I \citep{che17c}. LSQ14mo does not show any signs of strong \cii. However, perhaps interestingly LSQ14mo displays \oii-line morphology similar to SN~2018bsz.
A full analysis of how SN~2018bsz fits into the overall picture of SLSNe-I will be possible when the full post-peak
evolution has been observed.

\section{SN~2018bsz photometric properties}

\begin{figure*}
\centering
\includegraphics[width=17cm]{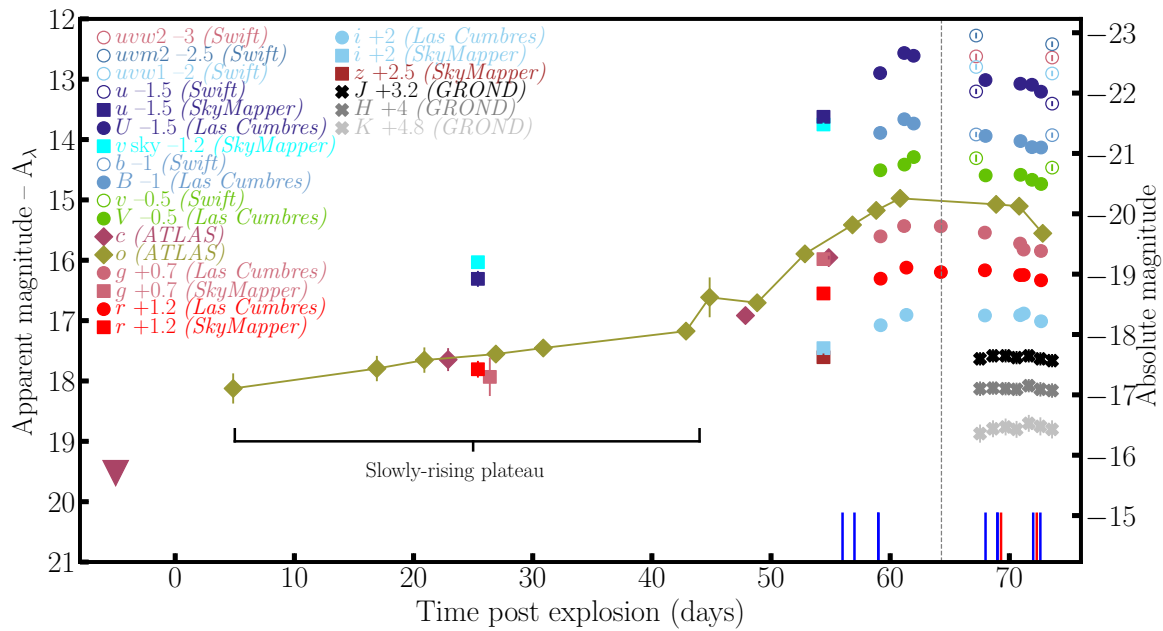}
\caption{UV, optical and near-IR light curves of SN~2018bsz. 
All photometry is corrected for both Milky Way and host galaxy line of sight extinction.
Epochs of optical spectroscopy are shown
as blue lines on the x-axis, while epochs of near-IR spectroscopy are indicated by red lines.
The upper limit of the most recent non-detection prior to discovery is shown as a maroon upside-down triangle.}
\label{lcatlas}
\end{figure*}

\subsection{Photometric data}
UV, optical and near-IR photometry of SN~2018bsz were obtained from a range of different telescopes
and instruments. Photometry for each instrument is listed in the tables in the appendix. ATLAS photometry is measured from point-spread-function fitting 
through \texttt{tphot} \citep{mer15} applied to the difference images. The orange
and cyan filters and the photometric calibration  are described in 
\cite{ton18}, 
and the magnitudes are on the AB system. ATLAS typically observes
each region of sky 4 times per night in 30\,s exposures over a $\sim$1\,hr 
period and we report the nightly mean magnitude here (Table~\ref{atlasphot}).\\
\indent Near UV photometry was obtained using the Neil Gehrels \textit{Swift} Observatory
 on 2018 May 31, and 2018 June 3. The transient was bright in all near-UV (NUV) and optical filters of the \textit{Swift}/UVOT telescope. The UVOT data were
 reduced using the standard pipeline available in the HEAsoft
 software package\footnote{https://heasarc.nasa.gov/lheasoft/}. The observation on May 31 was conducted during 3 orbits while the
 observation on June 3 was conducted during 2 orbits. To improve the
 S/N ratio of the observation in a given band in a particular epoch
 we co-added all orbit-data for that corresponding epoch using
 the HEAsoft routine \texttt{uvotimsum}. We used the routine \texttt{uvotdetect} to
 determine the correct position\footnote{This position is fully consistent with the updated transient coordinates provided by ATLAS that are stated in Section 2.} of the transient and used the routine
 \texttt{uvotsource} to measure the apparent magnitude of the transient by performing
 aperture photometry. For source extraction we
 used an aperture with a radius 4.7$''$ while to determine the
 background an aperture of radius 127.8$''$ was used. Magnitudes listed in Table~\ref{swiftphot} are on the AB system. We note that the UVOT photometry is not host subtracted.\\
\indent 
We monitored the $JHK$ light curve evolution of SN~2018bsz using the Gamma-Ray Burst Optical/Near-Infrared
Detector (GROND; \citealt{gre08}), mounted at the 2.2\,m MPG telescope at the ESO La Silla Observatory in Chile. The images were reduced using the GROND pipeline \citep{kru08},
which applies de-bias and flat-field corrections, stacks images and provides astrometry calibration. 
The SN magnitudes are calibrated against 2MASS field stars and are listed in Table~\ref{groundphot} in the AB system.\\ 
\indent Las Cumbres Observatory $UBVgri$-band data were obtained with Sinistro cameras on the 1m telescopes, through the Global Supernova Project. Using lcogtsnpipe \citep{val16}, a \texttt{PyRAF}-based photometric reduction pipeline, PSF fitting was performed. $UBV$-band data were calibrated to Vega magnitudes \citep{ste00} using standard fields observed on the same night by the same telescope. $gri$-band data were calibrated to AB magnitudes using the the AAVSO Photometric All-Sky Survey (APASS, \citealt{hen09}). Sinistro
photometry is listed in Table~\ref{LCOphot}. It is important to note that the Sinistro photometry is not host-galaxy subtracted due to a lack of host templates.\\
\indent SkyMapper photometry was extracted from images from the Transient Survey \citep{sca17} and the Southern Survey \citep{wol18} taken using the set of $uvgriz$ filters available at the telescope. Images were reduced using the pipeline described in \cite{wol18}, where the photometric extraction and calibration methodology is also outlined. Magnitudes are obtained in the AB system, where zeropoint calibration is anchored to APASS DR9 and 2MASS. We note that the SkyMapper photometry is not host subtracted.

\subsection{The SN~2018bsz light curve, colours, and temperatures}
\begin{figure}
\centering
\includegraphics[width=\columnwidth]{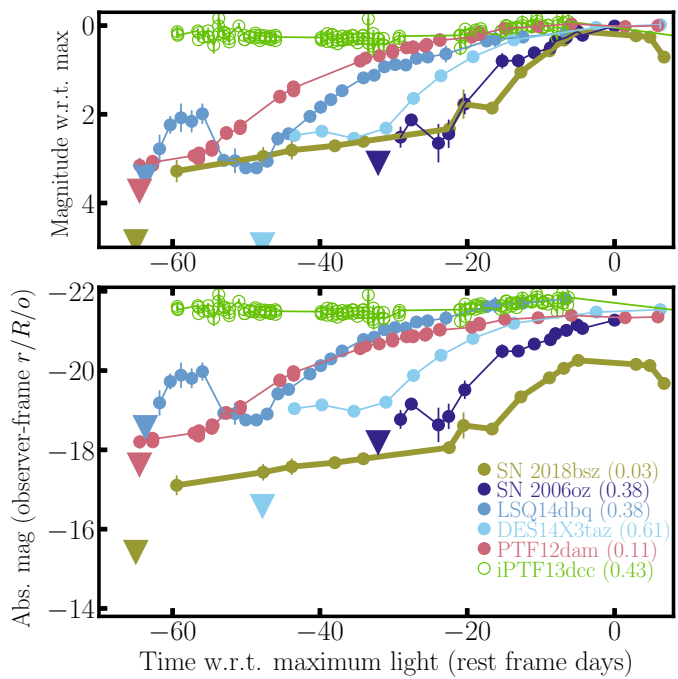}
\caption{Comparison of the pre-maximum light curve of SN~2018bsz to those of SLSNe-I from the literature showing clear bumps before the main rise to maximum (iPTF13dcc is shown in open circles as its form is clearly distinct from the events). Next to each SLSN name in parenthesis we give the object redshift (in the bottom panel).
Non-detection upper limits are indicated by upside down triangles. 
\textit{Top:} light-curves normalised to maximum-light magnitude.
\textit{Bottom:} absolute magnitude light curves with respect to maximum light. 
SN~2018bsz is in the ATLAS $o$ band, while all other comparison SNe are in
observed $R$ or $r$ band. Given the range of redshifts of the SLSN comparison sample, these observed bands equate to different rest frame wavelengths. However, given the lack of a) multi-band photometry (to enable rest-frame band comparison) and b) spectra of SLSNe-I during these epochs (to allow for valid $K$-corrections) we prefer to plot the data in its `raw' form. Colour effects may complicate the direct comparison of these light curves.}
\label{lccompSLSN}
\end{figure}

The NUV, optical and near-IR light curves of SN~2018bsz are displayed in Fig.~\ref{lcatlas}. 
To produce the absolute-magnitude light curves the data were corrected for a 
host galaxy distance modulus of 35.23, a Milky-Way reddening of E($B$-$V$)=0.214 mag \citep{Sch11}, and an E($B$-$V$)=0.041 mag internal to the host (a bolometric light curve is presented in Section 5).
The first detection gives an initial rise of 1.5 mag in five days (assuming an explosion epoch of 2018 March 25). 
The light curve then flattens, and a clear slowly rising `plateau' is observed from five to 33 days post explosion, equating to a duration of 26 days, that could easily be extended to around 40 days depending on which
photometric data one includes in the assessment.
During this plateau SN~2018bsz displays a relatively slow rise of 0.89 $o$-band magnitudes in 26 days (at a rate of 0.034\,mag\,d$^{-1}$), with an absolute magnitude of around --18 at around 25 days post explosion (40 days pre-maximum). A significant change in the light-curve shape is observed at just after 40 days,
with the rise significantly steepening. The light curve increases in brightness by an additional two magnitudes in 18 days (at a rate of 0.12\,mag\,d$^{-1}$). At peak (MJD = 58267.5, estimated from fitting a low-order polynomial to the $r$-band photometry) SN~2018bsz 
is --20.3 mag in the ATLAS $o$ band. SN~2018bsz has an absolute peak magnitude of --20.5 in the $B$ band and --20.3 in the $r$ band.\\
\indent SN~2018bsz thus has a relatively low absolute magnitude. Formally, this is significantly lower than the initial operational limit of --21 mag (at optical wavelengths) used to include SNe in the super-luminous category \citep{qui11,gal12a}. However, as outlined in the introduction, SLSNe are now more generally defined by spectroscopic properties, and while SN~2018bsz is somewhat spectroscopically abnormal compared to the general SLSN-I population (Section 3.2), it clearly shows similarities to SLSNe-I, as suggested by the SNID matches. The sample analyses of \cite{ins18b}, \cite{lun18}, and \cite{dec18} include a number of events dimmer than --21 mag, and SN~2018bsz falls within the published distributions while being a relatively low-luminosity SLSN-I. We also note that there exists a number of very bright SNe~Ic that are generally not considered SLSNe. Two examples are SN~2012aa \citep{roy16} and
SN~2011kl \citep{gre15} that both peaked at around --20 mag. Understanding the link between luminous SNe~Ic and under-luminous SLSNe-I may give important clues to the physics of these different classes of explosions.
\textit{Swift} photometry three days post maximum gives an absolute \textit{uvw1} (central wavelength of 2600\,\AA) magnitude of --20.4. Comparison to figure 6 of \cite{smi18a} suggests that SN~2018bsz is offset to lower luminosity than the rest of the sample, by almost one magnitude. The post-maximum near-IR light curve is relatively flat with little change in brightness over the $\sim$1-week duration of the observations.\\
\indent Using the estimated epoch of maximum brightness in the $r$-band gives a rise time of 63 rest-frame days. This is quite typical for SLSNe-I as shown in Fig.~\ref{lccompSLSN} and is close to the representative value used in \cite{dec18}. However, a number of SLSNe-I with both significantly longer and shorter rise times have been observed.
The latest photometry displayed in Fig.~\ref{lcatlas} indicates that SN~2018bsz is now clearly past maximum light and has started to decline. A characterisation of the post-maximum light curve awaits further data.\\
\indent We estimate the observed colours (corrected for Milky Way and host galaxy extinction) of SN~2018bsz at four epochs: during the plateau; at the start of the second steeper rise; at maximum light (in the $r$ band); and from our last photometric data obtained at eight days post maximum.
At around 25 days post explosion/40 days before maximum SkyMapper multi-band photometry is available. This gives a $u-g$ colour of 0.58 mag at the mid-point of the initial slower rise/plateau (see Fig.~\ref{lcatlas}). 
At around 55 days post explosion or ten days before maximum (when the light curve has clearly started to steepen its rise), the $u-g$ colour becomes much bluer, at --0.16 mag. SN~2018bsz continues to get bluer with a $U-g$ colour (note here we are using Las Cumbres Observatory photometry and no longer SkyMapper) at maximum light of --0.54. The last epoch of photometry included  in this paper, at eight days post maximum, shows that the SN has started to get redder again with a $U-g$ colour of --0.44 mag.\\
\indent To estimate black-body temperatures we extract spectral energy distributions (SEDs) from the NUV and optical photometry of SN~2018bsz at the same epochs for which we cite colours in the previous paragraph. Temperature errors are estimated through combining those obtained through Monte Carlo simulations sampling values from the photometry and their errors together with those obtained by removing the bluest and reddest photometric bands and reestimating temperatures.
At the earliest epoch where we have more than two photometric points -- at around 25 days post explosion or 40 days before maximum -- using SkyMapper $uvgr$ photometry we estimate relatively cool black-body temperature of 6700$\pm$1000\,K. This rises to 13,300$\pm$940\,K after the light curve has started the main rise to maximum (at around 50 days post explosion or 10 days before maximum, again estimated using SkyMapper photometry but now $uvgriz$). At around maximum light a temperature of 9500$\pm$180K is obtained (now using Las Cumbres photometry; $UBVgri$), while at eight days post maximum the temperature cools to 8800$\pm$200\,K (again using Las Cumbres $UBVgri$ photometry).\\ 
\indent The estimated temperature changes are consistent with the colour changes outlined above. Black-body fitting to spectra produces similar results. Overall, we observe that SN~2018bsz was quite red and cool during it's long-duration plateau, but quickly became blue and hot as the rise steepened to maximum light.
However, we note that the SkyMapper and Las Cumbres photometry have not been subtracted of the underlying host emission, and this is probably affecting the estimated colours. Indeed, ATLAS photometry -- that is host-galaxy subtracted --  gives $c-o$\,$\approx$\,0 at almost the same plateau epoch of 25 days post explosion. This is bluer than the colour from SkyMapper, and implies a temperature in excess of 9000\,K. In addition, the ATLAS colour does not change significantly from the plateau to the start of the rise to peak, in contrast to the SkyMapper and Las Cumbres colours.
Therefore the above colours, black-body temperatures and their time evolution should be evaluated with some caution and a full discussion of these parameters for SN~2018bsz will only be possible when all presented SN photometry is host-galaxy subtracted in the future.
At the same time, it seems unlikely that the temperature during the plateau is much higher than 10,000\,K, and this suggests that overall SN~2018bsz is relatively red at these epochs, while during the main rise it also appear to be likely that SN~2018bsz does not reach the high temperatures of 15-20,000\,K reached by other SLSNe-I \citep{how17,bos18}.  

\begin{figure}
\centering
\includegraphics[width=\columnwidth]{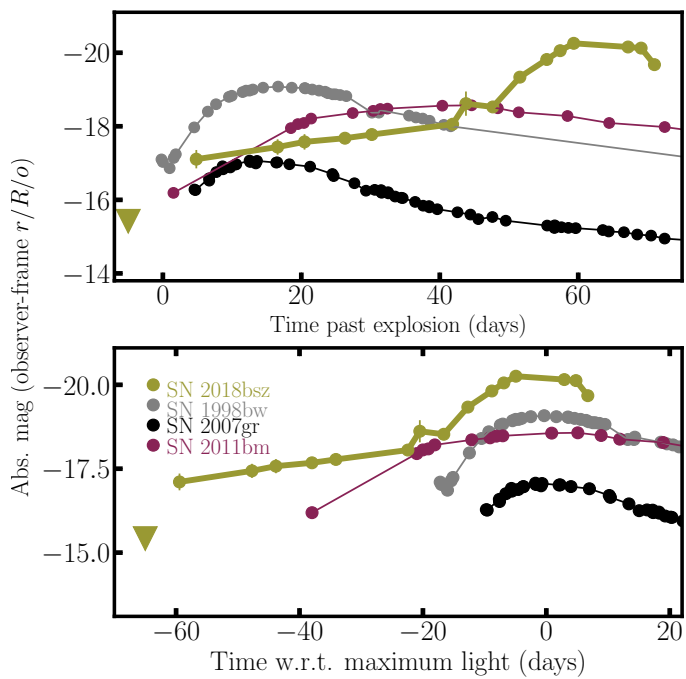}
\caption{Comparison of light curve of SN~2018bsz to SNe~Ic: SN~1998bw, a broad-line event associated with a long-duration Gamma-Ray Burst; SN~2007gr, a typical SN~Ic; and SN~2011bm, a slowly evolving SN~Ic.
\textit{Top:} light curves with respect to estimated explosion epoch.
\textit{Bottom:} light curves with respect to epoch of maximum light}
\label{lccompSNIc}
\end{figure}

\subsection{The pre-maximum light curve of SN~2018bsz compared to other SLSNe-I}
As discussed above, SN~2018bsz shows a clear, long-duration rising plateau before the second steeper rise to maximum light. Pre-rise `bumps' or `excess emission' have previously been observed in a number
of SLSNe-I \citep{lel12,nic15a,nic16a,smi16,vre17,lun18}. However, the feature observed in SN~2018bsz is not well described by the term bump as it has a linear form (in magnitudes) for at least 26 days (and up to 40 days) before the steeper
rise to maximum starts. As shown in the previous section, the colour of SN~2018bsz during this initial epoch is relatively red implying a relatively low black-body temperature.\\
\indent Figure ~\ref{lccompSLSN} compares ATLAS photometry of SN~2018bsz to that of five
other SLSNe from the literature that showed clear signs of an early-time bump before the main rise to maximum light. (We note, the photometry from these comparison sample are not $K$-corrected, leading to a strong caveat to some of the subsequent discussion: all comparison SNe are shown in observed-frame $r$ or $R$ band.) There are two properties
that mark SN~2018bsz as being different from these previously discussed bumps. First, the feature in SN~2018bsz appears to be of significantly longer duration. Taking the minimum length of the observed plateau of SN~2018bsz to be 26 days, this is then around ten days longer than any of the comparison sample. Secondly, the absolute magnitude at the mid-point of this plateau is around --18 mag (in ATLAS $o$ band).\\ 
\indent Most other events are significantly brighter than SN~2018bsz (both at this epoch but also at maximum; see also events in \citealt{lun18}), with only PTF12dam \citep{vre17} displaying similar luminosity pre-rise emission. However, in the case of PTF12dam, the strong constraint on the explosion epoch suggests that this bump was only of around 10 days in duration (while the upper limit plotted in Fig.~\ref{lccompSLSN} is only 0.5 mag fainter than the first detection, a non-detection three days before has an upper limit 1.8 mag fainter, hence ruling out an unobserved longer-duration plateau).\\
\indent SN~2006oz \citep{lel12} shows a light-curve morphology somewhat similar to SN~2018bsz (see top panel of Fig.~\ref{lccompSLSN}), but brighter by around a magnitude both at maximum and during the plateau (bottom panel of Fig.~\ref{lccompSLSN}). The upper-limit on the non detection of SN~2006oz prior to discovery is only 0.6 mag lower than the first detection. This suggests the possibility that the bump in SN~2006oz could have extended further back in time but was below our detection limits. However, \cite{lel12} constrained the temperature to be around 15,000\,K during these early epochs; higher than that of SN~2018bsz.\\
\indent The case of LSQ14dbq \citep{nic15a} is clearly distinct, as this SN displayed a very pronounced first peak of around 10 days duration before declining then rising again to the second brighter maximum. Finally, DES14X3taz \citep{smi16} displays a somewhat similar light-curve morphology to SN~2018bsz, while being more than a magnitude brighter. However, in this case the upper limit prior to the start of this bump rules out a more extended plateau as observed in SN~2018bsz. In addition, similarly to the case of SN~2006oz: DES14X3taz was constrained to have a significantly higher temperature during this bump.\\ 
\indent While we plot iPTF13dcc \citep{vre17} in Fig.~\ref{lccompSLSN} (given that it is often discussed within this context in the literature), we do not discuss it any further here, given that the light-curve morphology is quite distinct from the other events shown.
Overall, we conclude that the relatively red, long-duration slow rise (plateau) seen in SN~2018bsz is unprecedented in SLSNe-I observed to date.\\
\indent Aware of the clear uniqueness of the early-time light curve of SN~2018bsz, we now briefly discuss previous models suggested to explain bumps in the early-time light curves of other SLSNe (a full exploration of possible models to explain the peculiar properties of SN~2018bsz is beyond the scope of this initial discovery paper).
One option is that the initial plateau is simply the initial `normal' core-collapse SN of for example type Ic (to match the later carbon-dominated spectra). In the cases of LSQ14bdq and DES14X3taz this possibility was ruled out due to the high luminosity and very large $^{56}$Ni mass required \citep{nic15a,smi16}. SN~2018bsz is significantly dimmer during the early-time light curve, at a luminosity similar to some SNe~Ibc \citep{ric14}.\\
\indent In Fig.~\ref{lccompSNIc} we plot three example SNe~Ic. SN~1998bw was a broad-line SN~Ic associated with a Gamma-Ray Burst (data obtained from \citealt{pat01}), SN~2007gr was a normal SN~Ic \citep{val08}, and 
SN~2011bm was a slowly evolving SN~Ic \citep{val12}. We plot these SNe~Ic both with respect to epoch of maximum (as done for all SLSNe-I in the figure) and with respect to explosion epoch. The latter is done to specifically see how well a SN~Ic light curve may be able to explain the early-time pre-maximum light curve of SN~2018bsz. SN~1998bw is significantly brighter than SN~2018bsz at 10-20 days after explosion, and the morphology of the light curves are also distinct. 
SN~2007gr has a similar luminosity 10-15 days after explosion, but quickly declines while SN~2018bsz continues to rise. Meanwhile, SN~2011bm matches the pre-maximum light curve of SN~2018bsz reasonably well. The latter is somewhat flatter and shows no clear peak before the rise steepens to maximum, however the differences between SN~2018bsz and SN~2011bm are not so big.\\
\indent We estimate the black-body temperature of SN~2011bm from the available multi-band photometry as achieved for SN~2018bsz above at a similar epoch (post explosion) to the plateau temperature of SN~2018bsz.
In the case of SN~2011bm we obtain a temperature of 7200$\pm$650\,K that is consistent with that of SN~2018bsz at a similar epoch (6700$\pm$1000\,K, although note the above caveats). Hence, we suggest the possibility that the plateau in SN~2018bsz could have been produced by a SN~Ic-like explosion, where the following steeper brightening is due to a secondary (possibly central) power source that starts to dominate at later times. In the case of SN~2011bm, \cite{val12} estimated a massive ejecta and therefore a very large initial progenitor mass from the broad light curve.
As an additional note of interest Fig.~\ref{lccompSNIc} (bottom panel) shows that with respect to maximum light epoch the light curve of SN~1998bw has a very similar morphology to that of SN~2018bsz only at lower luminosity.\\
\indent Previously, a more favoured model for other SLSNe-I showing signs of excess emission pre-maximum has been that of shock cooling of
extended material around the progenitor (see \citealt{lel12,nic15a,smi16} and the analytic model of \citealt{pir15}). In the case of SN~2018bsz, shock-cooling has to be extended to much longer durations and produce a slowly rising light curve. It is not clear that shock cooling can produce such a light curve.
\cite{kas16} suggested that a magnetar-driven shock breakout could produce double-peaked light curves such as that observed in LSQ14dbq. However, these models predict temperatures of around 20,000\,K during the first peak; much higher than the colour/temperature observed in SN~2018bsz making the \cite{kas16} scenario unlikely.

\section{Comparison to synthetic spectra}
\begin{figure}
\centering
\includegraphics[width=\columnwidth]{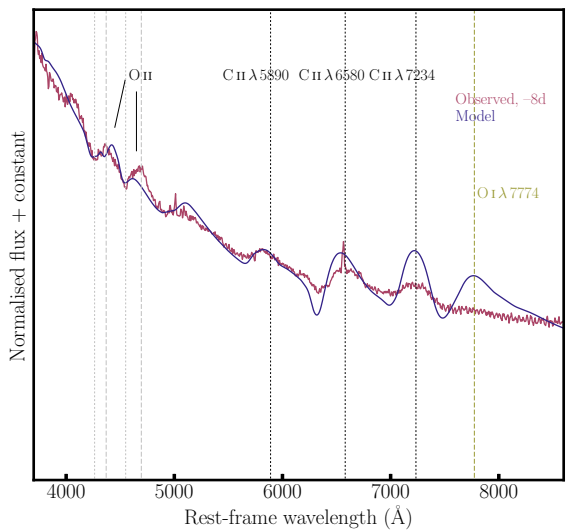}
\caption{Comparison of an optical-wavelength spectrum of SN~2018bsz 
with a model spectrum from Dessart (in preparation).
\cii\ line rest-frame wavelengths are indicated by black dotted vertical lines. In the case of the \oii\ lines we draw vertical grey dashed lines that align with the peaks in the spectra and vertical grey dotted lines for the troughs (absorption). \oi\,$\lambda$\,7774 is clearly observed in the model, but not in the observations, and is indicated by the vertical yellow dashed line.
(The model spectrum is at around 40 days post explosion and around 10 days post maximum and therefore does not correspond to the same exact epoch of the observed spectrum of SN~2018bsz.) }
\label{modelsluc}
\end{figure}

\begin{figure}
\centering
\includegraphics[width=\columnwidth]{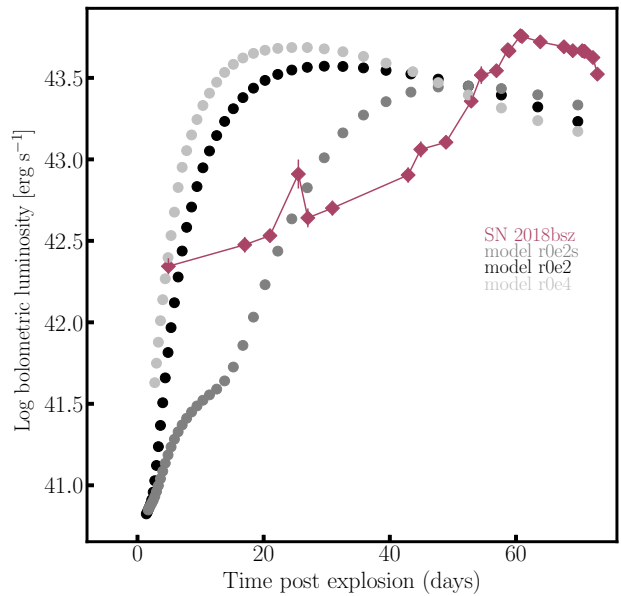}
\caption{Comparison of model bolometric light curves with that of SN~2018bsz. Model r0e2 is that discussed at length in the text. Model r0e2s is identical but with the magnetar power injected deep within the ejecta with little advection to large velocities, while r0e4 is identical to r0e2 but with higher kinetic energy. There appears to be a jump in the observed light curve at around 25 days post explosion. This is the epoch where multi-band photometry from SkyMapper is available and we see that SN~2018bsz is significantly brighter in $r$ (than other bands). This may be an issue with host galaxy contamination, and we will investigate this further when we are able to perform accurate galaxy-template subtraction when the SN has faded.}
\label{modelLC}
\end{figure}

The reclassification of SN~2018bsz to a SLSN-I was partially driven by comparison to 
model spectra, specifically the identification of strong \cii\ lines as outlined above.
In Fig.~\ref{modelsluc} we show a comparison between the observed spectrum obtained eight days before maximum light and a model spectrum that qualitatively shows similar line features. (The model spectrum is taken at 43 days post explosion, which corresponds to more than a week post maximum: later we discuss the clear differences between the model and SN~2018bsz after discussing the similarities in the spectra.)
This spectrum was taken from radiative transfer simulations from a theoretical
study on magnetar powered SNe of Type Ic (Dessart, in preparation).\\
\indent The numerical approach for these simulations is similar to that outlined in \cite{des18b}. 
Here, the progenitor we use is hydrogen-deficient. It corresponds to a 
carbon-rich Wolf-Rayet (WR) star with a final mass of 11.4\,\msun\ at core collapse
(model r0 of \citealt{des17b}, which had a progenitor of solar metallicity).
This progenitor star is exploded, with V1D \citep{liv93,des10c,des10a}
by means of a piston, producing a kinetic energy of $1.23 \times 10^{52}$\,erg. 
The kinetic energy is large relative to neutrino-powered SNe but is in line with what magneto-rotational explosions may produce (before forming a magnetar; \citealt{bur07,des08b}).
It contains (in part) the additional kinetic energy -- needed to match SLSN ejecta velocities -- induced by the magnetar (see below), and ignored by CMFGEN \citep{hil12} (this kinetic energy also matches those used in the spectral modelling of \citealt{maz16}).
The ejecta composition is dominated by oxygen, with a total oxygen mass of 5.59\,\msun, and only 0.2\,\msun\
of residual He in the outer ejecta -- the He surface mass fraction is 0.19.\\
\indent At one day after explosion, we remap this ejecta into CMFGEN
and proceed as for the GRB/SN models of \cite{des17b}.
To make this model deviate drastically from standard SNe Ic, we introduce 
a magnetar power $\dot{e}_{\rm pm}$.
For the magnetar properties, the model is characterised by a magnetic field of $3.5 \times 10^{14}$\,G
and an initial rotation energy of $4.0 \times 10^{50}$\,erg (corresponding to
an initial spin period of 7\,ms). The spin-down time scale for this model is 19.1 days,
the rise time is 25.8 days, and the peak luminosity is 
$4.86 \times 10^{43}$\,erg\,s$^{-1}$.
The energy deposition is prescribed rather than solved
for \citep{des18b}. Depending on the energy deposition profile, the steep light curve rise caused by 
magnetar power injection can occur within 2-3 days after explosion, or can be delayed by up to 20 days (Dessart, in preparation, Fig.~\ref{modelLC}).\\ 
\indent Compared to standard SNe~Ic, the influence of the magnetar is to raise the luminosity. Because the initial
rotational energy of the magnetar is small compared to the ejecta kinetic energy, the magnetar
power changes mostly the internal energy and escaping radiation. The peak luminosity is
10 times greater than in the model without a magnetar. This change is mostly carried by the
change in temperature in the spectrum formation region, which also causes a rise in ionisation.
The entire SED changes from red (SN~Ic) to blue, and the lines we see around maximum are those
of \oii\ and \cii. For \oii, the contributions arise from numerous multiplets around 4000\,\AA,
while the \cii\ lines arise from doublets at 5890\,\AA, 6580\,\AA, and triplets at 7235\,\AA.
\hei\,$\lambda$\,5875 contributes negligibly but \oi\,$\lambda$\,7774 is clearly present in the model.\\
\indent Compared to the observations the model spectrum qualitatively reproduces the main observed features at optical wavelengths (Fig.~\ref{modelsluc}), which we associate with \cii\ and \oii.
However, we note an offset in wavelength of the peaks of the \oii\ features between the observations and model.
As discussed in Section 3.2, this inconsistency of the spectral features seen in SN~2018bsz with observations of other SLSNe-I and now spectral models suggests differences in the properties of the
line forming regions in SN~2018bsz that should be the focus of future work.
\oi\,$\lambda$\,7774 is over predicted by the model. The observed near-IR spectra are almost featureless while the model displays \hei\,$\lambda$\,10830 (that we tentatively identify in observations, Fig.~\ref{nearIRseq}) together with a 
strong feature at around 9000\,\AA\ most likely associated with either \oi\ or \ciii\ (that is not observed).
The strong \cii\ lines observed in SN~2018bsz were also predicted by \cite{des12}, and are usually overestimated relative
to observations. Here, it is clear that SN~2018bsz has stronger
\cii\ lines than typically seen in SLSNe-I (see Fig.~\ref{spesnid}). In our models, the strong carbon lines stem from the
large carbon mass fraction in the WR progenitor model, either in the present model (computed with MESA; \citealt{pax11,pax13,pax15}), or the one from \citeauthor{des12} (2012, computed with KEPLER; \citealt{woo02}) -- both are from a 40\,\msun\ star.
It will be important for future work to investigate what controls 
the carbon/oxygen abundance in carbon-rich WR stars and whether this could be used to constrain the
evolutionary properties of the progenitor stars.\\
\indent It is important to note here that the models we use were not tailored to fit SN~2018bsz nor any other particular SN. Firstly, they were produced by exploding progenitors that were the result of standard stellar evolution (in this case using MESA) with the aim of using massive progenitors depleted in both hydrogen and helium (to match spectra of SLSN-I and SNe~Ic where the lack of these lines is one of their defining features). Secondly, the exact parameters used to produce models with different properties were chosen to be exploratory; to probe what produces differences in predicted observables. Changes in for example: ejecta mass; magnetar energy deposition; kinetic energy, can significantly change the model results. With these caveats in mind, in Fig.~\ref{modelLC} we plot three model bolometric light curves as compared to the observed bolometric light curve of SN~2018bsz (calculated using all available photometry and colours at all epochs, and following the methodology described in \citealt{ins18a}).\\
\indent All models have the same composition and are exploded in the same manner as discussed above, however compared to the model above (r0e2) in model r0e2s the magnetar power deposition profile is more confined in velocity space, while model r0e4 is identical to r0e2 but with higher kinetic energy.
It is clear that none of the model light curves reproduce that of SN~2018bsz. Most obvious is the long, slow (and bright) rise of SN~2018bsz that is missing in the models, plus the later epoch of maximum light. However, model r0e2s, where the magnetar power is much more confined to the inner ejecta is informative: it shows that a simple change in the way the magnetar energy is deposited can have significant effects on the shape of the light curve. This suggests that fine tuning of the chosen progenitor, explosion and magnetar properties could be able to come closer to the observed properties of SN~2018bsz. One possibility is that this slow rise is indicative of a more massive ejecta with a deeply embedded power source. Of course, there may also be multiple power sources at play producing the observed features of SN~2018bsz.\\
\indent The model comparison presented in this section shows that the explosion of a carbon-rich WR progenitor can produce spectra qualitatively similar to SN~2018bsz under the correct ionisation conditions.
While the timing of these model conditions does not match that SN~2018bsz (in terms of the light-curve evolution), the comparison allows for the identification of strong, persistent \cii\ lines in this nearby SLSN. Understanding what drives the differences between the models and SN~2018bsz can further aid in our understanding of SLSNe.

\section{Host galaxy properties}
\begin{figure}
\includegraphics[width=\columnwidth]{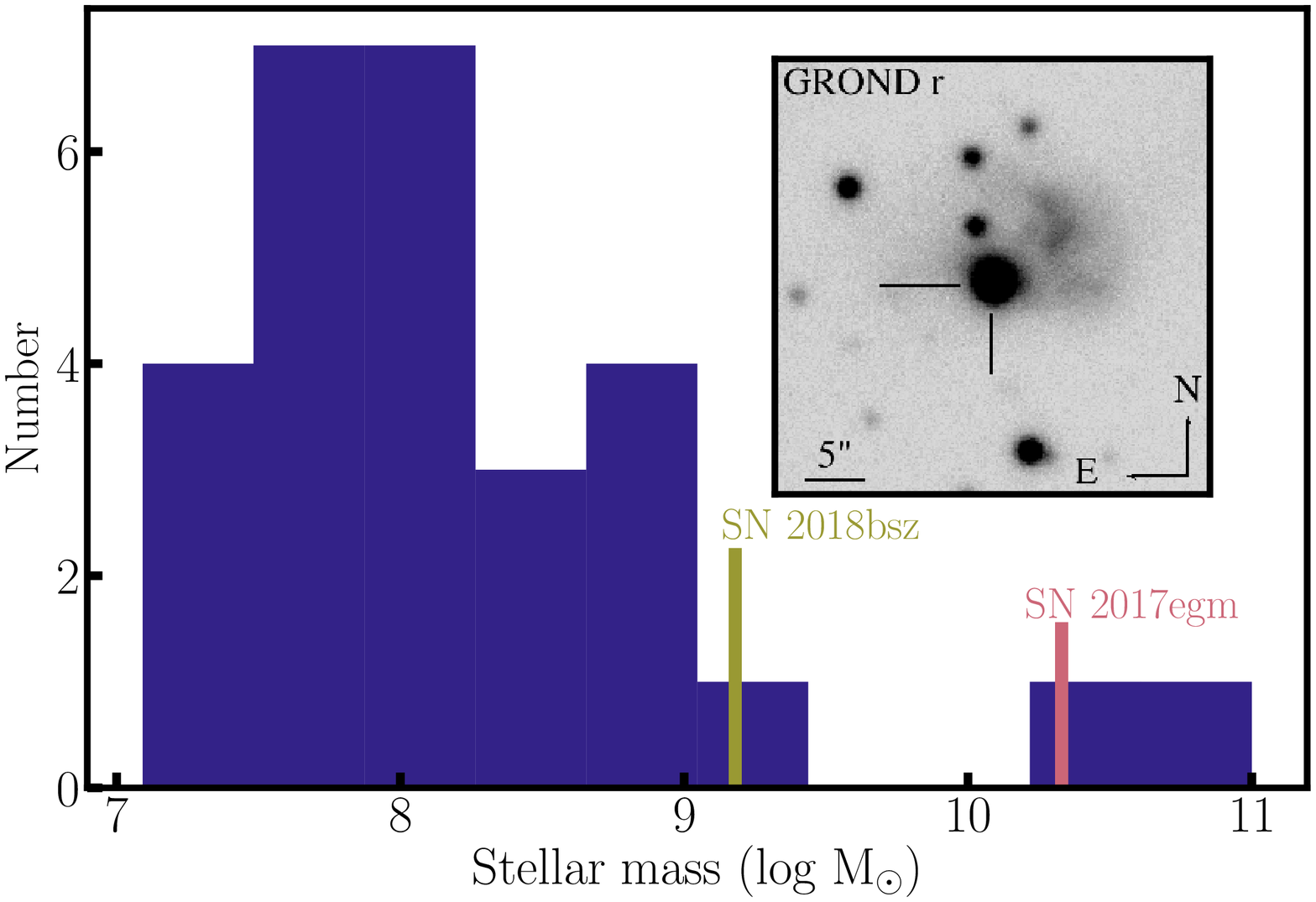}
\includegraphics[width=\columnwidth]{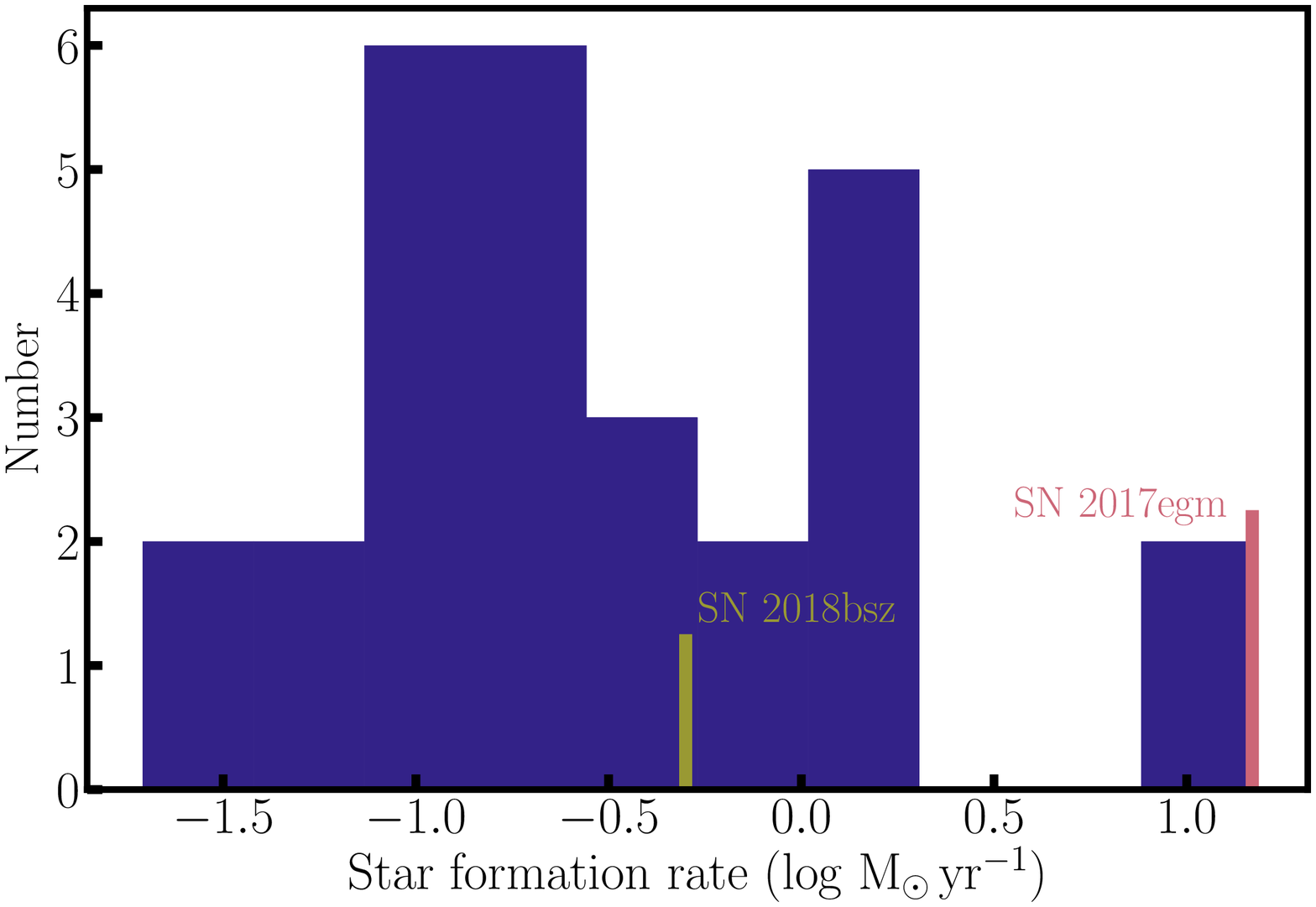}
\includegraphics[width=\columnwidth]{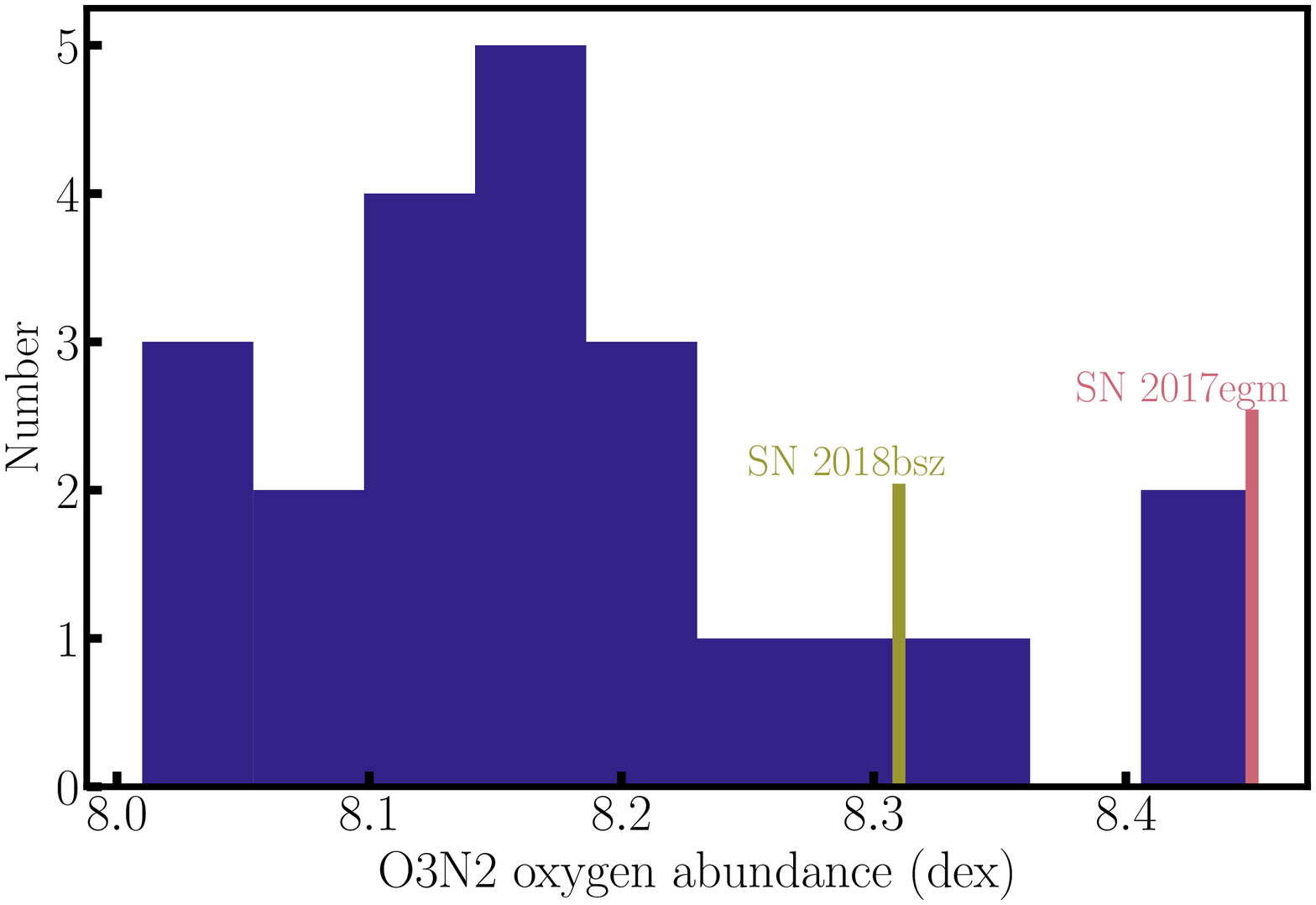}
\caption{Histograms of the environment and host properties of SN~2018bsz as compared
to literature samples and SN~2017egm; the previously lowest-redshift SLSN. \textit{Top:} integrated host galaxy stellar mass. The inset shows a GROND $r$-band image of SN~2018bsz (indicated by the two perpendicular black lines) and its host galaxy. \textit{Middle:} integrated host SFR. \textit{Bottom:} host environment oxygen abundance on the O3N2 scale (a combination of global galaxy and nearby host \hii\ region measurements). On each plot the value of SN~2018bsz is indicated, together with that of SN~2017egm.}
\label{hostprops}
\end{figure}

The explosion environments of transients can give clues to their progenitor properties, by constraining their probable ages and metallicities (see \citealt{and15b} for a review).
Early in the exploration of SLSNe, it was suggested that they prefer dwarf hosts \citep{nei11}, and this (together with their intrinsic low rate) is often cited (see e.g. \citealt{lel15}) as a reason that they went undetected until around a decade ago, given the previous bias in SN searches to massive, nearby galaxies.
Several subsequent studies have analysed samples of SLSNe-I, concluding that their host galaxy properties are biased towards dwarf galaxies of high specific star-formation rate (sSFR) and low, significantly sub-solar metallicity as compared to other SN types and the general galaxy population \citep{lun14,lel15,per16,che17a,sch18}. These studies have suggested that low metallicity may be a prerequisite for the production of SLSN events, perhaps in addition 
to a young age \citep{lel15,sch18}, putting
significant constraints on any proposed progenitor model. However, the nearby (the nearest until SN~2018bsz) SLSN-I, SN~2017egm, occurred in a host of relatively high mass and around solar metallicity challenging this picture (\citealt{bos18,nic17,che17b}, although see \citealt{izz18}). It is in this context 
that we analyse the host galaxy of SN~2018bsz.\\
\indent The host galaxy system (2MASX J16093905-3203443) is well-detected and resolved in our supernova follow-up imaging (top panel of Fig.~\ref{hostprops}).  The SN is coincident with what appears to be an isolated spiral dwarf galaxy (17$\arcsec$ or 9 kpc in diameter) with a disturbed morphology, at an offset of 6.8$\arcsec$ from the diffuse nucleus.  However, pre-imaging from Pan-STARRS (Fig.~\ref{diffimg}) makes clear that there is an additional galaxy directly under the SN position concealed by the SN light in the post-explosion imaging (and not distinct in shallow publicly available imaging catalogues, below), suggesting that the system is in fact an ongoing merger of two dwarf galaxies. These properties are reminiscent of many previous SLSN hosts (\citealt{lun14,per16,che17c}). Detailed characterisation of the host system of SN~2018bsz will be the focus of future work.\\ 
\indent Additional insight into the nature of the system can be obtained by SED modelling of the multi-wavelength galaxy continuum. The two galaxies are not cleanly separable in pre-explosion imaging at most wavelengths, so we model them together as a single system.  Data from the UV to the near-IR were obtained from the Galaxy Evolution Explorer \citep[\textit{GALEX};][]{mar05}, the Two Micron All-Sky Survey \citep[2MASS;][]{huc12} and the Wide-field Infrared Survey Explorer \citep[WISE;][]{wri10} public databases.  Skymapper does not report catalogue photometry for extended objects, so we performed our own photometry on the SkyMapper images using an aperture radius of 29 pixels (14.4$\arcsec$; 7.7 kpc), establishing the photometric calibration by direct comparison to nearby stars in the SkyMapper catalogue \citep{wol18}.\\
\indent We then fit the broad-band SED of 2MASX J16093905-3203443 using  \verb Le   \verb Phare  \citep{ilb06} to determine
the stellar mass and star-formation rate as in Taggart et al. (in preparation). The results indicate 2MASX J16093905-3203443 is a star-forming, low-mass galaxy with a moderate amount of dust. 
The galaxy has an absolute $r$-band magnitude of --19.8$\pm$0.24: somewhat brighter than the Large Magellanic Cloud (LMC) but still relatively faint compared to the host population of typical core-collapse supernovae. The stellar mass is M$_{*}$ = 1.5$^{+0.08}_{-0.33}$ $\times$10$^{9}$ M$_{\odot}$.  The star-formation rate is constrained only poorly; we derive SFR = 0.50$^{+2.22}_{-0.19}$ M$_{\odot}$ yr$^{-1}$. \\
\indent We also extract an \hii\ region spectrum from a 2-dimensional SN~2018bsz spectrum in order to
derive an environment oxygen abundance. The EFOSC2 spectrum taken at --6 days was used and an emission-line spectrum was extracted from the peak of the galaxy emission (in the 2-dimensional spectrum), at a distance of 3.3\,kpc from the SN. Emission-line fluxes were then measured
for all \hii\ region lines detected and are listed in Table~\ref{tablines}.
Using the O3N2 diagnostic on the \cite{mar13} scale we estimate an oxygen abundance of 8.31$\pm$0.01 dex (this error is that arising from the measurement error of the line fluxes; the \citealt{mar13} O3N2 diagnostic gives an additional 0.18 dex systematic error), consistent with around half solar.\\
\indent In Fig.~\ref{hostprops} the environment oxygen abundance, host stellar mass, and host star-formation rate are displayed as compared to samples of SLSNe in the nearby Universe. To create these comparison samples we searched the literature for all SLSNe-I at a redshift lower than 0.3 with published host masses, SFRs and emission line fluxes (with the latter being taken from the exact SN sites if available and global host values if not). Emission-line fluxes were taken from \cite{lel15,per16,che17c,ins17}; and \cite{izz18} and oxygen abundances were calculated on the \cite{mar13} O3N2 scale. Host masses and SFRs were estimated using
the samples published in \cite{lel15,per16,sch18}, however the PTF hosts presented in \cite{per16} were remodelled using \verb Le   \verb Phare. Within these distributions we also plot the position of each parameter for SN~2018bsz and SN~2017egm (Taggart et al. in preparation).\\ 
\indent As discussed above, SN~2017egm was the previously lowest-redshift SLSN-I and was an outlier in terms of its host galaxy properties, being more massive and more metal rich than the majority of previously observed events.  \cite{bos18} argued that the absence of similarly low-redshift SLSNe-I in low-metallicity hosts lowered the importance of progenitor metallicity in the production of these events.   In contrast, Fig.~\ref{hostprops} shows that the host of SN~2018bsz is quite typical of the previously-known SLSN-I host population.  While its mass and oxygen abundance are towards the upper end of the distribution from the comparison samples, it is not exceptional.  We argue that the small sample of very low-$z$ SLSN-I hosts (two events at $z<0.05$) is consistent with our present understanding of the SLSN-I host population based on intermediate-redshift studies.

\section{Conclusions}
We have presented early-time UV, optical and near-IR data of the lowest-redshift SLSN-I 
discovered to date, SN~2018bsz. Photometry from the ATLAS survey shows an unprecedented slowly rising (red), long-duration plateau before the main SN peak, with the former being distinct from all other previously analysed bumps in pre-maximum SLSN light curves. The spectra -- obtained only after the steeper rise to maximum light had started -- display uncharacteristically strong, persistent \cii\ features that are qualitatively similar to those produced by a magnetar-powered explosion of a massive WR progenitor. The host galaxy of SN~2018bsz does not stand out from the general SLSN-I host population, being an intermediate-mass, relatively low metallicity galaxy.\\ 
\indent Here, we have analysed  and discussed the properties of SN~2018bsz until around a week post maximum light. This nearby SLSN-I will still be visible until mid-October 2018 (before going behind the Sun), and therefore a full multi-wavelength characterisation of the post-peak evolution will be possible. When SN~2018bsz again becomes visible in late January 2019 nebular-phase spectroscopy will provide strong constraints on the progenitor star, its explosion, and the power source at the origin of this unique light curve. In conclusion, SN~2018bsz has already provided surprises as to the diversity and nature of SLSNe-I and further discoveries await the coming months.

\begin{acknowledgements}
We thank the staff at Paranal for their efficiency in obtaining our observations.
James Leftley is thanked for help with the SED fitting of the SN photometry.
This work is based (in part) on observations collected at the European Organisation for Astronomical 
Research in the Southern Hemisphere, Chile as part of ePESSTO, (the extended Public ESO Spectroscopic Survey for Transient Objects Survey) programme 199.D-0143.
Based on (in part) observations collected at the European Organisation for Astronomical Research in the Southern Hemisphere under ESO programme(s) 2101.D-5023(A).
Based on observations obtained at the Gemini Observatory under programme GS-2018A-Q-107 (PI: Sand). Gemini is operated by the Association of Universities for Research in Astronomy, Inc., under a cooperative agreement with the NSF on
behalf of the Gemini partnership: the NSF (United States), the National
Research Council (Canada), CONICYT (Chile), Ministerio de Ciencia, Tecnolog\'ia e Innovaci\'on Productiva (Argentina), and Minist\'erio da Ci\^encia, Tecnologia e Inova\c{c}\~ao (Brazil).
Part of the funding for GROND (both hardware as well as personnel) was generously granted from the Leibniz-Prize to Prof. G. Hasinger (DFG grant HA 1850/28-1). 
SJS acknowledges funding from STFC Grant Ref: ST/P000312/1. 
GL is supported by a research grant (19054) from VILLUM FONDEN.
Parts of this research were conducted by the Australian Research Council Centre of Excellence for All-sky Astrophysics (CAASTRO), through project number CE110001020. TWC acknowledgments the funding provided by the Alexander von Humboldt Foundation.
Support for this work was provided by NASA grant NN12AR55G.
Research by DJS is supported by NSF grants AST-1821987 and 1821967.
This work makes use of observations from the Las Cumbres Observatory network.  DAH, CM, and GH are supported by NSF grant AST 1313484.
A.G.-Y. is supported by the EU via ERC grant No. 725161, the Quantum Universe I-Core program, the ISF, the BSF Transformative program and by a Kimmel award.
Support for IA was provided by NASA through the Einstein Fellowship Program, grant PF6-170148. LG was supported in part by the US National Science Foundation under Grant AST-1311862.
MG is supported by the Polish National Science Centre grant OPUS
2015/17/B/ST9/03167.
MF is supported by a Royal Society - Science Foundation Ireland University Research Fellowship.
KM is supported by STFC through an Ernest Rutherford fellowship.
MB acknowledges support from the Swedish Research Council (Vetenskapsr\aa det) and the Swedish National Space Board.
AJR is funded by the Australian Research Council through grant number FT170100243. MDS is funded by a research grant (13261) from the Villum foundation. 
Parts of this project were conducted by the Australian Research Council Centre of Excellence for All-sky Astro-physics (CAASTRO), through project number CE110001020.
EYH and CA acknowledge the support provided by the National Science Foundation under grant No. AST-1613472.
ZKR acknowledges support from European Research Council Consolidator Grant 647208. 
This research was supported by the Australian Research Council Centre of Excellence for All Sky Astrophysics in 3 Dimensions (ASTRO 3D), through project number CE170100013. This research has made use of data, software and/or web tools obtained from the High Energy Astrophysics Science Archive Research Center (HEASARC), a service of the Astrophysics Science Division at NASA/GSFC and of the Smithsonian Astrophysical Observatory's High Energy Astrophysics Division.
This publication makes use of data products from the Two Micron All
Sky Survey, which is a joint project of the University of
Massachusetts and the Infrared Processing and Analysis
Center/California Institute of Technology, funded by the National
Aeronautics and Space Administration and the National Science
Foundation.
We thank the Swift ToO team for
executing of our observations. Pan-STARRS is supported by the University of Hawaii and the NASA’s Planetary Defense Office under Grant no. NNX14AM74G. 
GALEX (Galaxy Evolution Explorer) is a NASA Small Explorer, launched
in April 2003. We gratefully acknowledge NASA’s support for
construction, operation, and science analysis of the GALEX mission,
developed in cooperation with the Centre National d’Etudes Spatiales
of France and the Korean Ministry of Science and Technology.
This publication makes use of data products from the Wide-field
Infrared Survey Explorer, which is a joint project of the University
of California, Los Angeles, and the Jet Propulsion
Laboratory/California Institute of Technology, funded by the National
Aeronautics and Space Administration.
The national facility capability for SkyMapper has been funded through ARC LIEF grant LE130100104 from the Australian Research Council, awarded to the University of Sydney, the Australian National University, Swinburne University of Technology, the University of Queensland, the University of Western Australia, the University of Melbourne, Curtin University of Technology, Monash University and the Australian Astronomical Observatory. SkyMapper is owned and operated by The Australian National University's Research School of Astronomy and Astrophysics. The survey data were processed and provided by the SkyMapper Team at ANU. The SkyMapper node of the All-Sky Virtual Observatory (ASVO) is hosted at the National Computational Infrastructure (NCI). Development and support the SkyMapper node of the ASVO has been funded in part by Astronomy Australia Limited (AAL) and the Australian Government through the Commonwealth's Education Investment Fund (EIF) and National Collaborative Research Infrastructure Strategy (NCRIS), particularly the National eResearch Collaboration Tools and Resources (NeCTAR) and the Australian National Data Service Projects (ANDS).
\end{acknowledgements}

%-------------------------------------------------------------------
\bibliographystyle{aa}

\bibliography{Reference}
%\begin{thebibliography}{}

%\end{thebibliography}

\appendix
\section{SN~2018bsz spectral observations}
\begin{table*}
\centering
\begin{tabular}[t]{ccccccc}
\hline
Civilian date (UT) & MJD & Epoch$^{a}$ & Telescope & Instrument & Wavelength range (\AA) & Grism/Grating\\
\hline	
2018-05-20 & 58258.5 & --9  & FTN & FLOYDS & 3200--10000 & 235\,l/mm\\
2018-05-21 & 58259.2 & --8  & NTT & EFOSC2 & 3700--9300  & Gr13     \\
2018-05-23 & 58261.3 & --6  & NTT & EFOSC2 & 3400--7500  & Gr11     \\
2018-05-23 & 58261.3 & --6  & NTT & EFOSC2 & 6000--10000 & Gr16     \\
2018-05-23 & 58261.6 & --6  & FTS & FLOYD  & 3200--10000 & 235\,l/mm\\
2018-06-01 & 58270.0 &  +3  & LT  & SPRAT  & 4000--8000  & Wasatch600\\
2018-06-01 & 58270.3 &  +3  & VLT & X-Shooter & 3500--25000 & UVB+VIS+NIR\\
2018-06-01 & 58270.6 &  +3  & FTS & FLOYDS & 3200--10000 & 235\,l/mm\\
2018-06-05 & 58274.0 &  +7  & Magellan & IMACS & 4200--9400 & 300 l/mm blue\\
2018-06-05 & 58274.3 &  +7  & Magellan & FIRE & 8800--20000 & LDPrism\\
2018-06-05 & 58274.4 &  +7  & FTN & FLOYDS & 3200--10000 & 235\,l/mm\\
2018-06-06 & 58275.0 &  +8  & LT & SPRAT & 4000--8000 & Wasatch600\\
2018-06-08 & 58277.0 &  +10  & Gemini-South & FLAMINGOS-2 & 9900--18000 & $JH$ grism\\
\hline                      
\hline                      
\end{tabular}   
\caption{Log of SN~2018bsz spectral observations. $^{a}$With respect to the $r$-band maximum of MJD = 58267.5.}
\label{tabspec}
\end{table*}

\section{SN~2018bsz photometry}
\begin{table}
\centering
\begin{tabular}[t]{ccc}
\hline
MJD & band & Magnitude (error)\\
\hline                      
58207.5 &$o$&  18.73 (0.25)\\
58219.5 &$o$&  18.40 (0.21)\\
58223.5 &$o$&  18.26 (0.21)\\
58225.5 &$c$&  18.44 (0.19)\\
58229.5 &$o$&  18.16 (0.04)\\
58233.5 &$o$&  18.06 (0.06)\\
58245.5 &$o$&  17.78 (0.12)\\
58247.4 &$o$&  17.22 (0.33)\\
58250.5 &$c$&  17.71 (0.14)\\
58251.4 &$o$&  17.31 (0.09)\\
58255.4 &$o$&  16.50 (0.13)\\
58257.4 &$c$&  16.75 (0.03)\\
58259.4 &$o$&  16.02 (0.03)\\
58261.4 &$o$&  15.78 (0.02)\\
58263.4 &$o$&  15.58 (0.06)\\
58271.5 &$o$&  15.68 (0.03)\\
58273.4 &$o$&  15.71 (0.01)\\
58275.4 &$o$&  16.16 (0.02)\\
\hline                      
\hline                      
\end{tabular}   
\caption{ATLAS AB optical host-subtracted photometry. These photometry are nightly averages of (in general) four individual exposures.}
\label{atlasphot}
\end{table}

\begin{table}
\centering
\begin{tabular}[t]{ccc}
\hline
MJD & band & Magnitude (error)\\
\hline
58228.0 &$u$& 19.07 (0.13)\\
58228.0 &$v$& 18.41 (0.05)\\
58228.0 &$r$& 17.32 (0.14)\\
58229.0 &$g$& 18.12 (0.32)\\
58257.0 &$u$& 16.38 (0.03)\\
58257.0 &$v$& 16.13 (0.03)\\
58257.0 &$g$& 16.17 (0.05)\\
58257.0	&$r$& 16.07 (0.05)\\
58257.0 &$i$& 15.95 (0.05)\\
58257.0 &$z$& 15.98 (0.06)\\
\hline                      
\hline                      
\end{tabular}   
\caption{SkyMapper AB non-host subtracted photometry. Multiple exposures on any given night are averaged to give the values presented here.}
\label{skyphot}
\end{table}

\begin{table}
\centering
\begin{tabular}[t]{ccc}
\hline
MJD & band & Magnitude (error)\\
\hline
58261.8 &$U$& 15.65 (0.01)\\
58261.8 &$B$& 15.94 (0.01)\\   
58261.8 &$g$& 15.86 (0.05)\\
58261.8 &$V$& 15.80 (0.02)\\
58261.8 &$r$& 15.77 (0.01)\\
58261.8 &$i$& 15.57 (0.01)\\
58263.8 &$U$& 15.32 (0.03)\\
58263.8 &$B$& 15.71 (0.02)\\
58263.8 &$g$& 15.67 (0.04)\\
58263.8 &$V$& 15.71 (0.03)\\
58264.6 &$U$& 15.36 (0.01)\\
58264.6 &$B$& 15.77 (0.01)\\
58264.6 &$V$& 15.58 (0.02)\\
58264.0 &$r$& 15.59 (0.02)\\
58264.0 &$i$& 15.40 (0.02)\\
58266.8 &$g$& 15.69 (0.05)\\
58266.9 &$r$& 15.66 (0.04)\\
58270.6 &$U$& 15.77 (0.05)\\
58270.6 &$B$& 15.99 (0.01)\\
58270.5 &$g$& 15.80 (0.02)\\
58270.6 &$V$& 15.89 (0.03)\\
58270.5 &$r$& 15.63 (0.01)\\
58270.6 &$i$& 15.41 (0.01)\\
58273.5 &$U$& 15.83 (0.01)\\
58273.5 &$B$& 16.08 (0.01)\\
58273.5 &$g$& 15.98 (0.01)\\
58273.5 &$V$& 15.88 (0.01)\\
58273.8 &$r$& 15.71 (0.01)\\
58273.5 &$i$& 15.41 (0.01)\\
58274.5 &$U$& 15.84 (0.02)\\
58274.5 &$B$& 16.18 (0.01)\\
58274.5 &$V$& 15.96 (0.01)\\
58275.3 &$U$& 16.00 (0.03)\\
58275.3 &$B$& 16.18	(0.01)\\
58275.3 &$g$& 16.10 (0.01)\\
58275.3 &$V$& 16.03 (0.01)\\
58275.3 &$r$& 15.80 (0.01)\\
58275.3 &$i$& 15.51 (0.01)\\
\hline                      
\hline                      
\end{tabular}   
\caption{Las Cumbres Observatory photometry. $UBV$ photometry is in the Vega system, while $gri$ photometry is in the AB system. Las Cumbres photometry has not been host-galaxy subtracted.}
\label{LCOphot}
\end{table}

\begin{table}
\centering
\begin{tabular}[t]{ccc}
\hline
MJD & band & Magnitude (error)\\
\hline	
58269.8 &$w2$& 17.73 (0.05)\\
58269.8 &$m2$& 17.19 (0.05)\\ 
58269.8 &$w1$& 16.50 (0.05)\\
58269.8 &$u$&  15.97 (0.04)\\
58269.8 &$b$&  15.97 (0.04)\\
58269.8 &$v$&  15.61 (0.06)\\
58276.2 &$w2$& 17.75 (0.05)\\
58276.2 &$m2$& 17.33 (0.05)\\
58276.2 &$w1$& 16.61 (0.05)\\
58276.2 &$u$&  16.17 (0.04)\\
58276.2 &$b$&  15.98 (0.04)\\
58276.2 &$v$&  15.76 (0.06)\\
\hline                      
\hline                      
\end{tabular}  
\caption{\textit{Swift} photometry on the AB system. Photometry is not host-galaxy subtracted.}
\label{swiftphot}
\end{table}

\begin{table}
\centering
\begin{tabular}[t]{ccc}
\hline
MJD & band & Magnitude (error)\\
\hline	
58270.1 &$J$ & 14.64 (0.11)\\
58270.1 &$H$ & 14.26 (0.12)\\
58270.1 &$K$ & 14.16 (0.15)\\
58271.2 &$J$ & 14.59 (0.11)\\
58271.2 &$H$ & 14.25 (0.11)\\
58271.2 &$K$ & 14.08 (0.14)\\
58272.3 &$J$ & 14.59 (0.11)\\
58272.3 &$H$ & 14.26 (0.11)\\
58272.3 &$K$ & 14.05 (0.14)\\
58273.2 &$J$ & 14.62 (0.11)\\
58273.2 &$H$ & 14.27 (0.11)\\
58273.2 &$K$ & 14.09 (0.14)\\
58274.2 &$J$ & 14.59 (0.11)\\
58274.2 &$H$ & 14.21 (0.11)\\
58274.2 &$K$ & 13.99 (0.14)\\
58275.2 &$J$ & 14.64 (0.11)\\
58275.2 &$H$ & 14.27 (0.12)\\
58275.2 &$K$ & 14.04 (0.15)\\
58276.2 &$J$ & 14.67 (0.11)\\
58276.2 &$H$ & 14.29 (0.12)\\
58276.2 &$K$ & 14.09 (0.15)\\
\hline                      
\hline                      
\end{tabular}   
\caption{GROND near-IR photometry on the AB system. GROND photometry has not
been host-galaxy subtracted.}
\label{groundphot}
\end{table}

\section{Emission-line fluxes}
\begin{table}
\centering
\begin{tabular}[t]{cc}
\hline
Emission line & Flux (error)\\
\hline	
[\oii]\,$\lambda\lambda$\,3727,3729 & 44.1 (0.83)\\
\hb\,$\lambda$\,4861 & 20.4 (1.08)\\
\oiii\,$\lambda$\,4959 & 16.86 (0.96)\\
\oiii\,$\lambda$\,5007 & 36.18 (0.86)\\
\ha\,$\lambda$\,6563 & 91.08 (1.07)\\
\nii\,$\lambda$\,6584 & 14.38 (0.98)\\
\sii\,$\lambda\lambda$\,6717,6731 & 21.36 (3.00)\\
\hline                      
\hline                      
\end{tabular}   
\caption{Emission-line fluxes measured from an \hii-region spectrum extracted
from 2-dimensional spectrum of SN~2018bsz. Fluxes are in units of 10$^{-16}$\,erg\,s$^{-1}$\,cm$^{-2}$.}
\label{tablines}
\end{table}

\end{document}